\documentclass[twocolumn]{aastex631}

\usepackage{natbib}
\usepackage[version=4]{mhchem}

\newcommand{\RNum}[1]{\uppercase\expandafter{\romannumeral #1\relax}}

\DeclareSymbolFont{UPM}{U}{eur}{m}{n}
\DeclareMathSymbol{\umu}{0}{UPM}{"16}
\let\oldumu=\umu
\renewcommand\umu{\ifmmode\oldumu\else\math{\oldumu}\fi}
\newcommand\micro{\umu}

\let\microns \jmicron

\shorttitle{Constraining \ce{O2} and \ce{O3} using Grids}
\shortauthors{Latouf et al.}

\begin{document}
\title{Bayesian Analysis for Remote Biosignature Identification on exoEarths (BARBIE) \RNum{2}: Using Grid-Based Nested Sampling in Coronagraphy Observation Simulations for \ce{O2} and \ce{O3}}

\author[0000-0001-8079-1882]{Natasha Latouf}
\altaffiliation{NSF Graduate Research Fellow, 2415 Eisenhower Ave, Alexandria, VA 22314}
\affiliation{Department of Physics and Astronomy, George Mason University, 4400 University Drive MS 3F3, Fairfax, VA, 22030, USA}
\affiliation{NASA Goddard Space Flight Center, 8800 Greenbelt Road, Greenbelt, MD 20771, USA}
\affiliation{Sellers Exoplanents Environment Collaboration, 8800 Greenbelt Road, Greenbelt, MD 20771, USA}

\author[0000-0002-8119-3355]{Avi M. Mandell}
\affiliation{NASA Goddard Space Flight Center, 8800 Greenbelt Road, Greenbelt, MD 20771, USA}
\affiliation{Sellers Exoplanents Environment Collaboration, 8800 Greenbelt Road, Greenbelt, MD 20771, USA}

\author[0000-0002-2662-5776]{Geronimo L. Villanueva}
\affiliation{NASA Goddard Space Flight Center, 8800 Greenbelt Road, Greenbelt, MD 20771, USA}
\affiliation{Sellers Exoplanents Environment Collaboration, 8800 Greenbelt Road, Greenbelt, MD 20771, USA}

\author[0000-0002-9338-8600]{Michael D. Himes}
\affiliation{NASA Postdoctoral Program Fellow, NASA Goddard Space Flight Center, 8800 Greenbelt Road, Greenbelt, MD 20771, USA}

\author[0000-0001-7912-6519]{Michael Dane Moore}
\affiliation{NASA Goddard Space Flight Center, Greenbelt, MD, USA.}
\affiliation{Business Integra, Inc., Bethesda, MD, USA.}
\affiliation{Sellers Exoplanets Environment Collaboration, 8800 Greenbelt Road, Greenbelt, MD 20771, USA}

\author{Nicholas Susemiehl}
\affiliation{Center for Research and Exploration in Space Science and Technology, NASA Goddard Space Flight Center, Greenbelt, MD, USA.}
\affiliation{NASA Goddard Space Flight Center, Greenbelt, MD, USA.}

\author[0000-0003-2273-8324]{Jaime Crouse}
\affiliation{NASA Goddard Space Flight Center, Greenbelt, MD, 20771}
\affiliation{NASA GSFC Sellers Exoplanet Environments Collaboration}

\author[0000-0003-0354-9325]{Shawn Domagal-Goldman}
\affiliation{NASA Goddard Space Flight Center, Greenbelt, MD, 20771}

\author[0000-0001-6285-267X]{Giada Arney}
\affiliation{NASA Goddard Space Flight Center, Greenbelt, MD, 20771}
\affiliation{NASA GSFC Sellers Exoplanet Environments Collaboration}

\author[0000-0002-5060-1993]{Vincent Kofman}
\affiliation{NASA Goddard Space Flight Center, 8800 Greenbelt Road, Greenbelt, MD 20771, USA}
\affiliation{Sellers Exoplanents Environment Collaboration, 8800 Greenbelt Road, Greenbelt, MD 20771, USA}
\affiliation{Integrated Space Science and Technology Institute, Department of Physics, American University, Washington DC}

\author[0000-0003-3099-1506]{Amber V. Young}
\affiliation{NASA Goddard Space Flight Center, Greenbelt, MD, 20771}
\affiliation{NASA GSFC Sellers Exoplanet Environments Collaboration}

\correspondingauthor{Natasha Latouf}
\email{nlatouf@gmu.edu, natasha.m.latouf@nasa.gov}

\begin{abstract}
We present the results for the detectability of the \ce{O2} and \ce{O3} molecular species in the atmosphere of an Earth-like planet using reflected light at visible wavelengths. By quantifying the detectability as a function of signal-to-noise ratio (SNR), we can constrain the best methods to detect these biosignatures with next-generation telescopes designed for high-contrast coronagraphy. Using 25 bandpasses between 0.515 and 1 {\microns}) and a pre-constructed grid of geometric albedo spectra, we examined the spectral sensitivity needed to detect these species for a range of molecular abundances. We first replicate a modern-Earth twin atmosphere to study the detectability of current \ce{O2} and \ce{O3} levels, and then expand to a wider range of literature-driven abundances for each molecule. We constrain the optimal 20\%, 30\%, and 40\% bandpasses based on the effective SNR of the data, and define the requirements for the possibility of simultaneous molecular detection. We present our findings of \ce{O2} and \ce{O3} detectability as functions of SNR, wavelength, and abundance, and discuss how to use these results for optimizing future instrument designs. We find that \ce{O2} is detectable between 0.64 and 0.83 {\microns} with moderate-SNR data for abundances near that of modern-Earth and greater, but undetectable for lower abundances consistent with a Proterozoic Earth. \ce{O3} is detectable only at very high SNR data in the case of modern-Earth abundances, however it is detectable at low-SNR data for higher \ce{O3} abundances that can occur from efficient abiotic \ce{O3} production mechanisms.
\end{abstract}


\section{Introduction}
\label{sec:intro}

Over 5,000 exoplanets have been discovered and confirmed in the last three decades. With the field of exoplanet exploration booming since the first exoplanet atmosphere discovery \citep[HD 209458;][]{charbonneau02, seager10, kaltenegger17, madhusudhan19}, the next hurdle is the detection and characterization of terrestrial exoplanet atmospheres. Current space-based observatories (e.g. Spitzer, HST, JWST) are beginning to probe the characteristics of potentially rocky planets with both transmission \citep{dewit2016,Lim2023} and emission \citep{kreidberg2019,zieba2023} measurements, but potentially habitable planets will be remain largely inaccessible except for 1-2 unique systems around the smallest stars (e.g. TRAPPIST-1) and impossible to characterize for Sun-like stars. However, with the advancements of high-contrast instrument technology and future mission concepts \citep[e.g.,][]{roberge18, luvoir, habex}, the ability to find a habitable Earth-twin around a Sun-like star is now a realistic goal for the next flagship visible-light space telescope, the Habitable Worlds Observatory (HWO). 

\begin{figure*}
\centering
\includegraphics[scale=0.5]{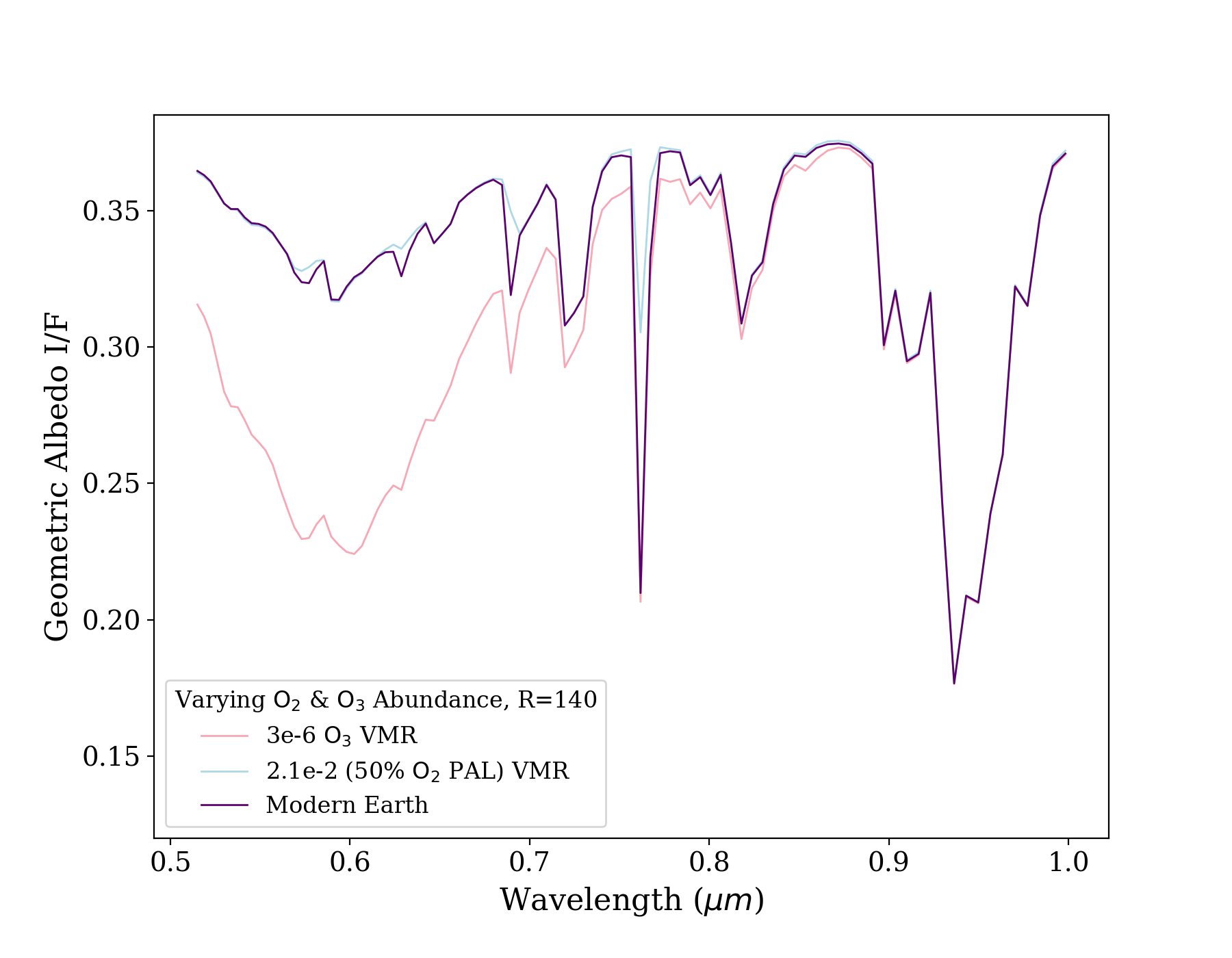}
\caption{Reflection spectra of a modern-Earth atmospheric composition but with varied \ce{O3} and \ce{O2} abundances. We feature a high \ce{O3} abundance (3$\times10^{-6}$), a 50\% PAL \ce{O2} abundance (2.1$\times10^{-2}$), and a modern Earth (7$\times10^{-7}$ \ce{O3}, and 2.1$\times10^{-1}$ \ce{O2}), respectively. All spectra are binned at R=140.}
\label{fig:example_abun}
\end{figure*}

In the search for signs of planetary habitability, the detection of potential biosignatures can hint at biological activity on the planetary surface. Since Earth is currently our only example of a conclusively habitable (and inhabited) planet, using model scenarios with conditions similar to those on modern Earth or its distant past serves as a starting point for examining the detectability of planetary characteristics \citep{arney16, rugheimer18}. In particular, at different times in its history Earth hosted varying concentrations of the atmospheric biomarkers \ce{O2} and \ce{O3}, which are primarily produced on Earth through photosynthesis and subsequent photochemical reactions; therefore measurements of geochemical proxies for the oxygenation of Earth's surface and atmosphere over time provide examples of possible global biogeochemical scenarios that we could encounter when observing Earth-like exoplanets. It is also useful to consider planets that - like Earth - have global liquid water oceans, but no biospheres. In some conditions, such 

The Phanerozoic (modern) epoch of Earth (541 million years ago -- present), which forms the basis for most habitability and biosignature studies, has relatively high levels of both \ce{O2} (21\%) and \ce{O3} (0.7 ppm) generated by abundant plant life and subsequent photochemistry. 
However, in the Proterozoic epoch (2.5 billion -- 541 million years ago), photosynthetic life was less abundant and measurements suggest oxygen accumulation in the atmosphere at approximately 0.1\% to 1\% of modern Earth \ce{O2} levels but with potentially detectable levels of \ce{O3} \citep{planavsky14}. 
In a planetary atmosphere similar to the Archean epoch of Earth (4 -- 2.5 billion years ago), we would expect an oxygen-poor and ozone-poor atmosphere, with a large abundance of greenhouse gases including \ce{CO2} and \ce{CH4}, which may have in turn formed a photochemical organic atmospheric haze \citep{zerkle12, arney16, krissansen18}. 
In addition to past epochs of Earth's history, models of alternative atmospheric scenarios are capable of producing of \ce{O2} and \ce{O3} dramatically higher than today's values - even without the presence of biology - due to a variety of mechanisms \citep{hu12, domagalgoldman14, gao15, tian15, harman18}.


As shown in Figure~\ref{fig:example_abun}, we can see how varying the abundances of \ce{O2} and \ce{O3} can alter the resultant visible spectra and lead to differences in the detectability of these gases. In this work we quantify the detectability of \ce{O2} and \ce{O3}, using reflected-light measurements at visible wavelengths (0.515 -- 1 {\microns}), for a range of abundance values and spectral bandpasses. This project is a direct continuation from \citet[][hereafter BARBIE1]{latouf23} - in BARBIE1, we studied the detection of \ce{H2O} as a function of SNR and abundance throughout the visible spectral range, 
and in this paper we extend the same methodology to the study of \ce{O2} and \ce{O3} and also examine the impact of the width of spectral bands on the SNR required for detection.
In $[\S]$ \ref{sec:method} we present the methodology of our simulations, also providing a brief summary of BARBIE1. In $[\S]$ \ref{sec:results} we present the results of our simulations for both the modern Earth-like SNR study and the molecular abundance study. In $[\S]$ \ref{sec:disc} we discuss the presented results and analyze the impact for future observations of varying Earth-twin epochs. In $[\S]$ \ref{sec:conc} we present our conclusions and ideas for future work.

\section{Methodology}
\label{sec:method}

We follow a similar methodological approach to that of BARBIE1. Herein we present a summary of the main steps in our analysis. 


\subsection{Inputs}
\label{inputs}

\subsubsection{Pre-Computed Spectral Grid}
\label{sec:grid}
We use a geometric albedo spectral grid that was pre-computed by \citet[][hereafter S23]{susemiehl23}. This grid is housed in the Planetary Spectrum Generator \citep[PSG,][]{PSG,PSGbook}. The grid contains 1.4 million geometric albedo spectra that have been pre-computed with the parameters, minimum, and maximum values laid out in Table 1 of S23. The native resolution of the grid is set to R=500 and binned to R=140 for our analysis. For more information on the creation and verification of the grid, see S23 and BARBIE1. There are three molecular species in the grid: \ce{H2O}, \ce{O3}, and \ce{O2}, with \ce{N2} as the assumed background gas. The minimum and maximum grid values for these parameters are as follows: \ce{H2O} in [$10^{-8}$, $10^{-1}$]; \ce{O3} in [$10^{-10}$, $10^{-1}$]; \ce{O2} in [$10^{-10}$, $0.8$]; \ce{N2} = 1 - \ce{H2O} - \ce{O3} - \ce{O2}. The model is comprised of 50\% clear and 50\% cloudy spectra, i.e. $\mathrm{C_{f}}$ = 50\%. There are three planetary parameters in the grid: surface pressure ($\mathrm{P_{0}}$), surface albedo ($\mathrm{A_{s}}$), and gravity (g). 
The minimum and maximum values for these parameters are as follows: $\mathrm{P_{0}}$ in [$10^{-3}$ Bar, 10 Bar]; $\mathrm{A_{s}}$ in [$10^{-2}$, 1]; gravity in [1 m/s$^{2}$, 100 m/s$^{2}$]; and $\mathrm{R_p}$ is fixed to 1 $R_\Earth$. 
The grid covers a wavelength range from 0.515 -- 1.0 {\microns} as this range has been defined as the VIS channel in the exoplanet imaging instrument concepts studied in the Astro2020 Decadal Survey that form the starting point for HWO. 


\begin{table}
    \centering
    \begin{minipage}{0.43\textwidth}
        \centering
        \begin{tabular}{cc}
            \hline
            \hline
            \textbf{Log-Step \ce{O2}} & \textbf{\ce{O2} (VMR)} \\
            \hline
            0 & 2.1$\times10^{-1}$$^{\dagger}$\\
            -0.25 & 1.2$\times10^{-1}$  \\
            -0.5 & 6.7$\times10^{-2}$\\
            -0.75 & 3.8$\times10^{-2}$\\
            -1 & 2.1$\times10^{-2}$ \\
            -1.25 & 1.2$\times10^{-2}$\\
            -1.5 & 6.7$\times10^{-3}$ \\
            -2 & 2.1$\times10^{-3}$ \\
            -2.5 & 6.7$\times10^{-4}$ \\
            -3 & 2.1$\times10^{-4}$\\
            \hline
        \end{tabular}
        \caption{The above values were used in our simulations for \ce{O2}, moving from Earth-like values into different epochs of Earth's history. \\$^{\dagger}$ Modern Earth-like value.}
        \label{tab:vals-o2}
    \end{minipage}%
    \hspace{0.1\textwidth} 
    \begin{minipage}{0.43\textwidth}
        \centering
        \begin{tabular}{cc}
            \hline
            \hline
            \textbf{Log-Step \ce{O3}} & \textbf{\ce{O3} (VMR)} \\
            \hline
            +0.65 & 3$\times10^{-6}$ \\
            +0.55 & 2.5$\times10^{-6}$ \\
            +0.45 & 2$\times10^{-6}$ \\
            +0.35 & 1.5$\times10^{-6}$ \\
            +0.25 & 1.25$\times10^{-6}$ \\
            +0.15 & 1$\times10^{-6}$ \\
            +0 & 7$\times10^{-7}$$^{\dagger}$ \\
            \hline
        \end{tabular}
        \caption{The above values were used in our simulations for \ce{O3}, moving from Earth-like values up to a high-\ce{O3} build-up regime. \\$^{\dagger}$ Modern Earth-like value.}
        \label{tab:vals-o3}
    \end{minipage}
\end{table}

\subsubsection{Mock Data and Retrieval Methodology}
\label{sec:fidspec}

Our fiducial ``data'' spectrum is primarily set as a modern-Earth twin following \citet{feng18}, with constant volume mixing ratios (VMRs) \ce{H2O}$=3\times10^{-3}$, \ce{O3}$=7\times10^{-7}$, \ce{O2}=$0.21$, a background gas of \ce{N2}=$1-$\ce{H2O}$-$\ce{O3}$-$\ce{O2}, constant temperature profile at 250 K, $\mathrm{A_{s}}$ of 0.3, $\mathrm{P_{0}}$ of 1 Bar, and a planetary radius fixed at $\mathrm{R_p}$ = 1 $\mathrm{R_\Earth}$. We consider a resolving power of 140, binned from the native grid resolving power of 500. This fiducial spectrum is given as data to the nested sampler in conjunction with the grid. We use the modern-Earth twin fiducial spectrum for our initial SNR study, wherein we focused on SNRs 3--16 moving in steps of 1. We then change the fiducial spectrum for our abundance study, changing only the molecule of interest one at a time, to the values listed in Table~\ref{tab:vals-o2} and ~\ref{tab:vals-o3}. All other parameters were left to modern-Earth values, in order to specifically constrain the molecule of interest. In BARBIE1, we focused on \ce{H2O}, centered on modern-Earth values and moving in increasing and decreasing log-steps. This was due to the lack of constraint on \ce{H2O} abundance throughout time. In this study, we consider \ce{O2} and \ce{O3}, and set our maximum \ce{O2} value as the modern-Earth value of 0.21 VMR and move in decreasing log-steps of 0.25 and 0.5 down to values of 1\% to 0.1\% to represent a mid-Proterozoic Earth epoch \citep{planavsky14}. For \ce{O3} we wished to test the possibility that values higher than on modern-day Earth might present stronger detectability. To study this potential, we set the maximum value as 3 $\times 10^{-6}$ VMR, a value that is $\sim5$x the modern-Earth value of 7 $\times 10^{-7}$ VMR, and well within the values created in models with high rates of abiotic \ce{O2} and \ce{O3} production on planets around F-, K, or M-type stars \citep{hu12, domagalgoldman14, gao15, tian15, harman18}. 
and decrease to the minimum value as the modern-Earth value.
In log space, we decreased in steps of 0.1, due to the small gap between the maximum and minimum values. 

As in BARBIE1, to examine the detectability of the molecular species as a function of the central wavelength and width of a bandpass,  we chose 25 evenly spaced values for the bandpass central wavelength. However, in this study we also vary the total width of the spectral bandpass, examining values of 20\%, 30\%, and 40\%. These ranges are consistent with the simultaneous bandpasses that may be achieved with future high-performance coronagraphs \citep{ruane15, por20, roser22}. Using these bandpass values and the inputs laid out in \ref{sec:grid} and \ref{sec:fidspec}, we run a series of Bayesian nested sampling retrievals for each abundance and SNR combination using the PSGnest application for the Planetary Spectrum Generator \citep[PSG;][]{PSG, PSGbook}.

PSG is a radiative transfer model and tool for synthesizing and retrieving upon planetary spectra. This includes planetary atmospheres and surfaces covering wavelengths from 50 nm to 100 mm (i.e. from UV to Radio) and a large range of planetary properties. PSG includes aerosol, atomic, continuum, and molecular scattering/radiative processes, implemented layer-by-layer. PSG also includes the nested sampling routine PSGnest \footnote{https://psg.gsfc.nasa.gov/apps/psgnest.php}, which is an adaptation of the algorithm used in Fortran MultiNest \citep{multinest}. PSGnest is a Bayesian retrieval tool based on the well-known MultiNest framework and designed for exoplanetary observations; for more information on PSGnest and our retrieval methodology, see S23 and BARBIE1.  

\subsection{Outputs}
\label{sec:outputs}

The output results file from PSGnest contains highest-likelihood values of output parameters, the average value resulting from the posterior distribution, uncertainties which are estimated from the standard deviation of the posterior distribution, and the log evidence (logZ) \citep{PSGbook}. It also includes the input data (wavelength, uncertainty, and input spectrum, in respective columns) with the best-fit spectrum. We calculate the median values, as well as the upper and lower limits of the 68\% credible region \citep{harrington22} using the output results. We also extract the posteriors and global log-evidence for use in our detectability calculations, which is the numerical representation that quantifies the relative support per each model given the input data \citep{multinest}. Using these outputs, we compute the Bayes factor. The Bayes factor is calculated by subtracting the Bayesian log-evidence per retrieval using the fiducial gas abundances. The resulting differences are referred to as the log-Bayes factor \citep[$\mathrm{lnB}$;][]{benneke13}. The log-Bayes factor determines which scenario is most likely by examining the hypothesis where all parameters are present, and systematically subtracting the evidences from scenarios where each parameter in turn is nullified. In our studies, if $\mathrm{lnB}$ is less than 2.5, it represents an unconstrained detection; if $\mathrm{lnB}$ is between 2.5 and 5, it is a weak detection; if $\mathrm{lnB}$ is greater than 5, it is a strong detection \citep[reference Table 2 of][]{benneke13}. This differs slightly from Table 2 of \citet{benneke13}, in which $\mathrm{lnB}$ represents a weak, moderate, and strong detection respectively. We do not calculate the log-Bayes factor for non-gaseous components, as those factors cannot be absent and thus the calculation cannot represent a detection of those components.

\begin{figure*}
\includegraphics[width=\textwidth]{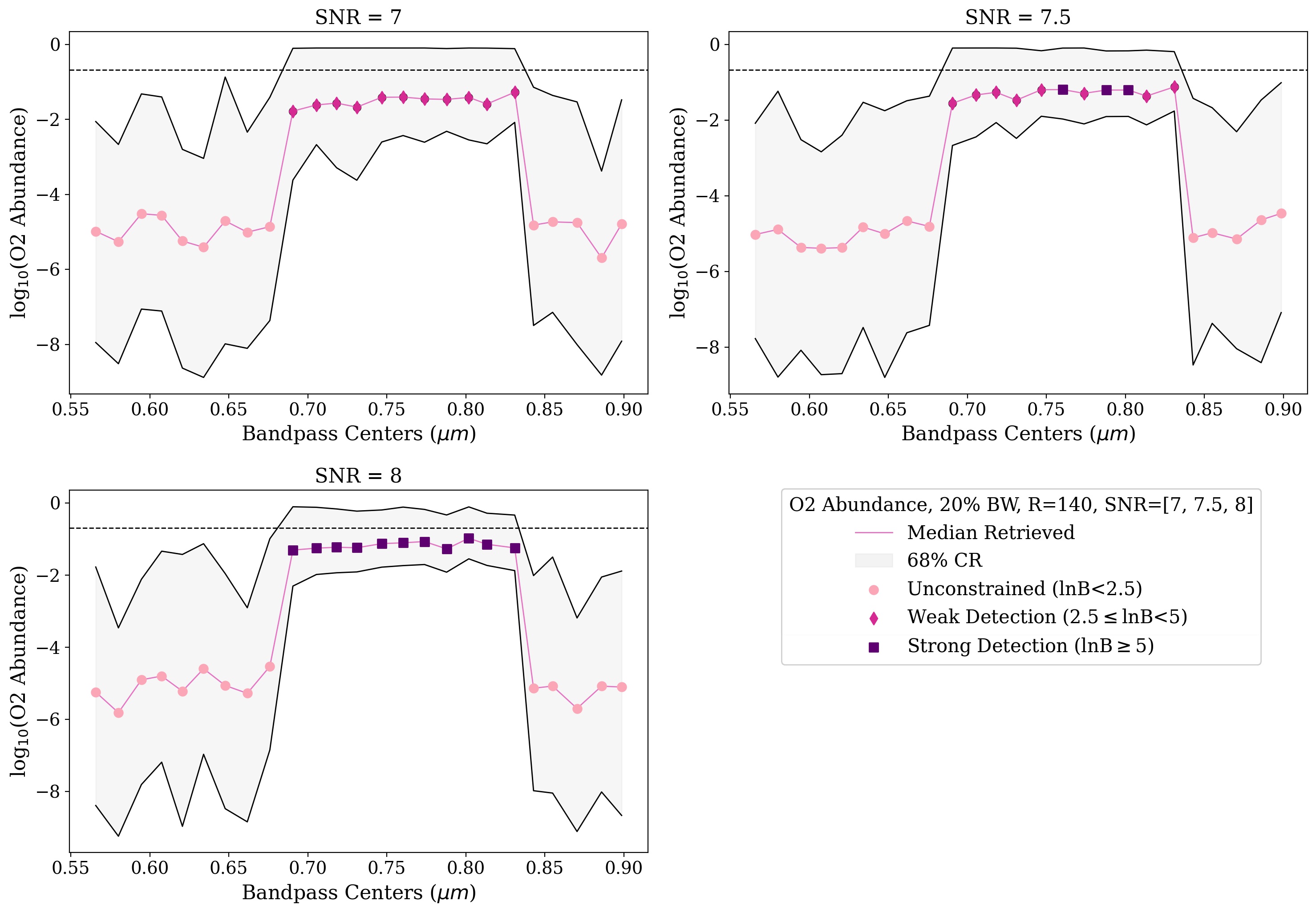}
\caption{Results of the fiducial case study for \ce{O2}, but showing only a narrow range of SNR values. The y-axis shows the abundance values for each molecule in log scale, with the true value is shown with the black dashed line (in this case, a modern Earth-like \ce{O2} value of 0.21 VMR). Each dot represents a bandpass center, with the pink line portraying the median retrieved values of \ce{O2} abundance for that bandpass, and the gray shaded region representing the upper and lower limits for the 68\% credible region. Increased retrieval certainty can be seen where the gray regions narrow. Each point is colored to indicate varying detection strength. Unconstrained regions ($lnB<2.5$) are shown in light pink dots, weak detections ($2.5 \leq lnB < 5$) are shown in dark pink diamonds, and strong detections ($lnB>5$) are shown in purple squares. We present the range at which detectability materially changes for \ce{O2}.}
\label{fig:o2_bfretest_narrow}
\end{figure*}

\begin{figure*}
\includegraphics[width=\textwidth]{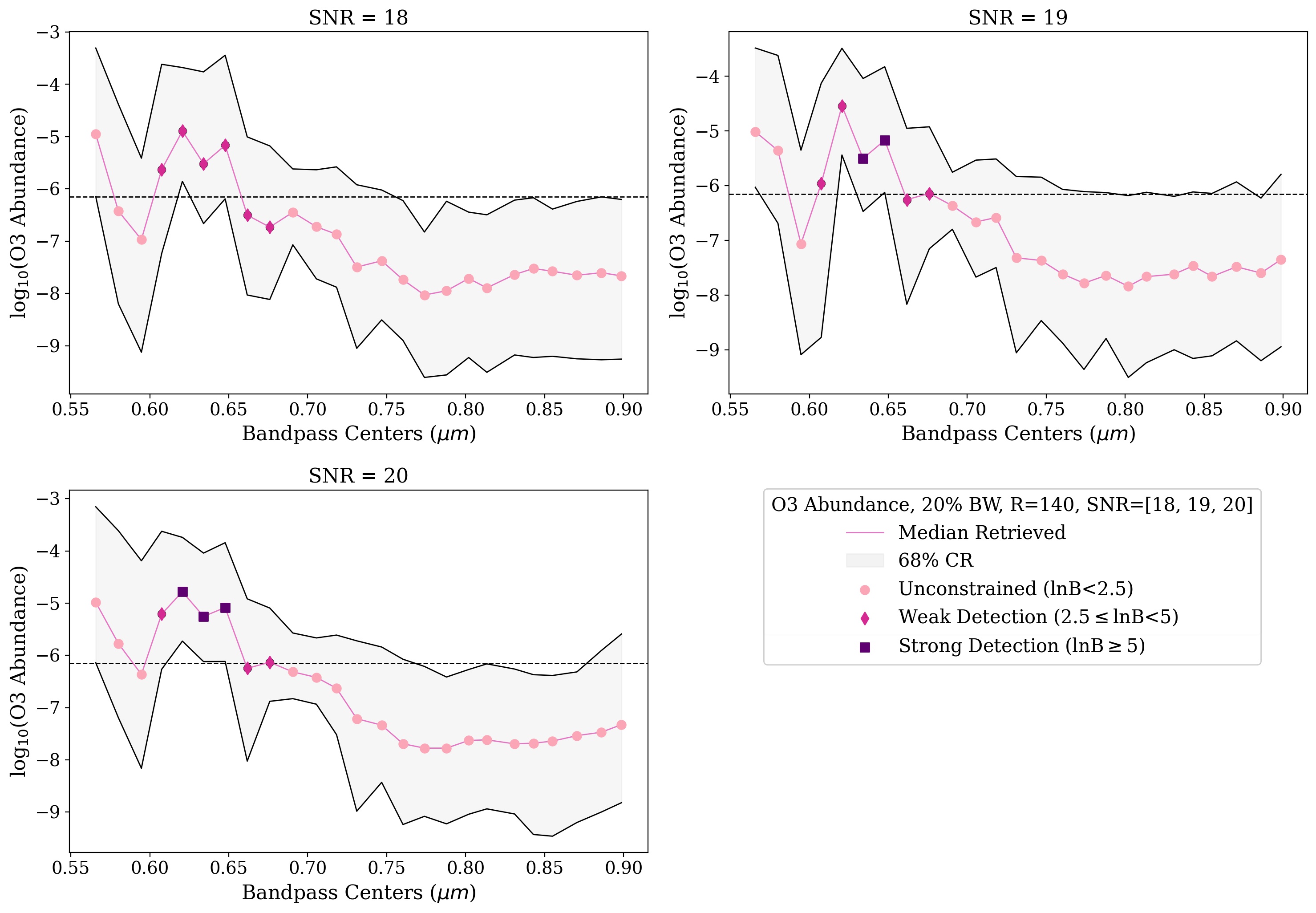}
\caption{Results of the fiducial case study for \ce{O3} but focused on a specific narrow SNR range. We show the SNR range at which detectability materially changes for \ce{O3}. All facets of the plot remain the same as in Figure~\ref{fig:o2_bfretest_narrow}. We notice that for SNR=20, 3 bandpasses achieve a strong detection but none of these bandpasses correctly constrain the abundance of \ce{O3} within the 68\% credible region. We present further investigation on this in Figure~\ref{fig:o3_corner}.}
\label{fig:o3_bfretest_narrow}
\end{figure*}

\begin{figure*}
\gridline{\fig{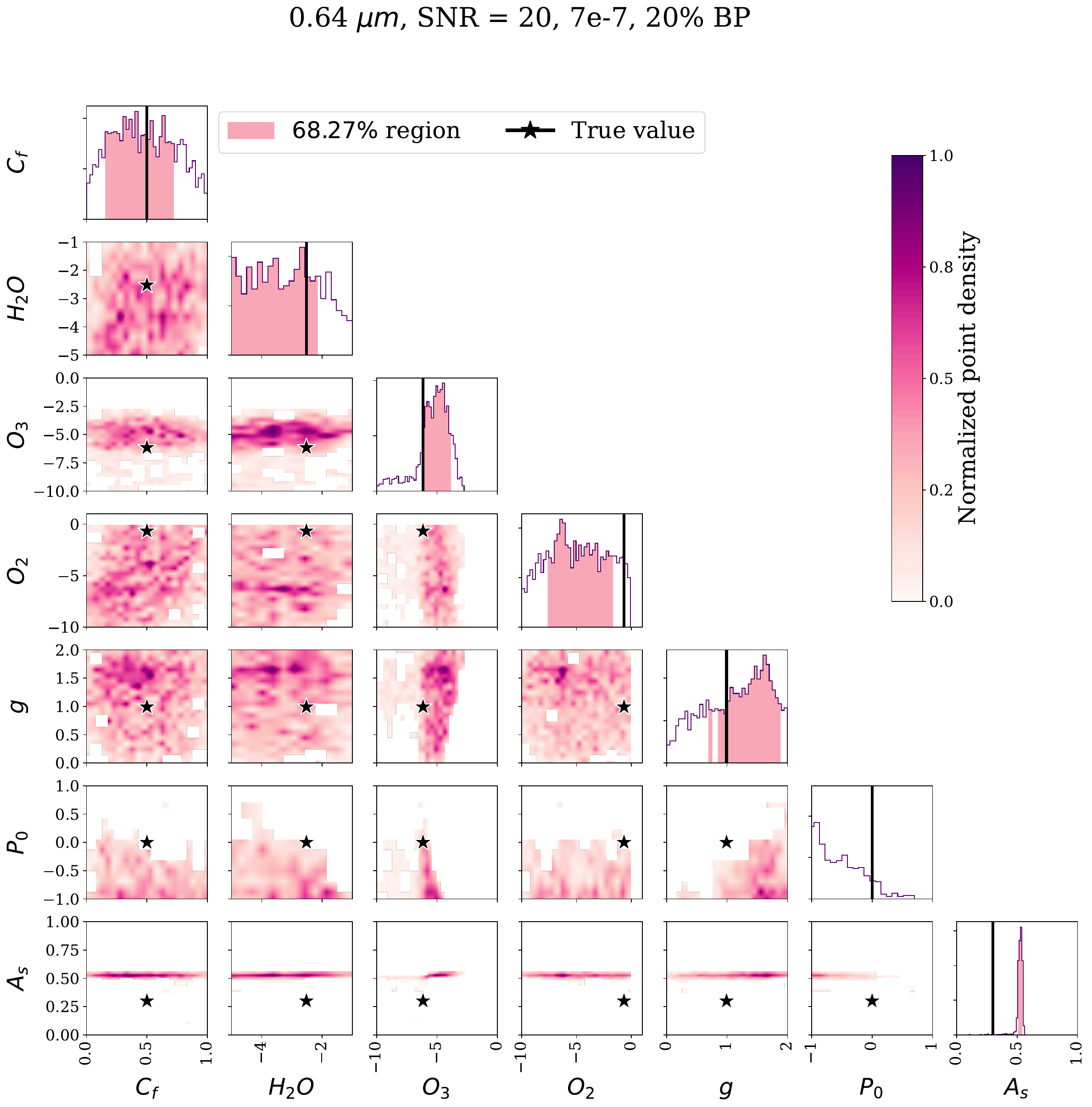}{0.5\textwidth}{(a) 
          20\% Bandpass, modern Earth values.}
          \fig{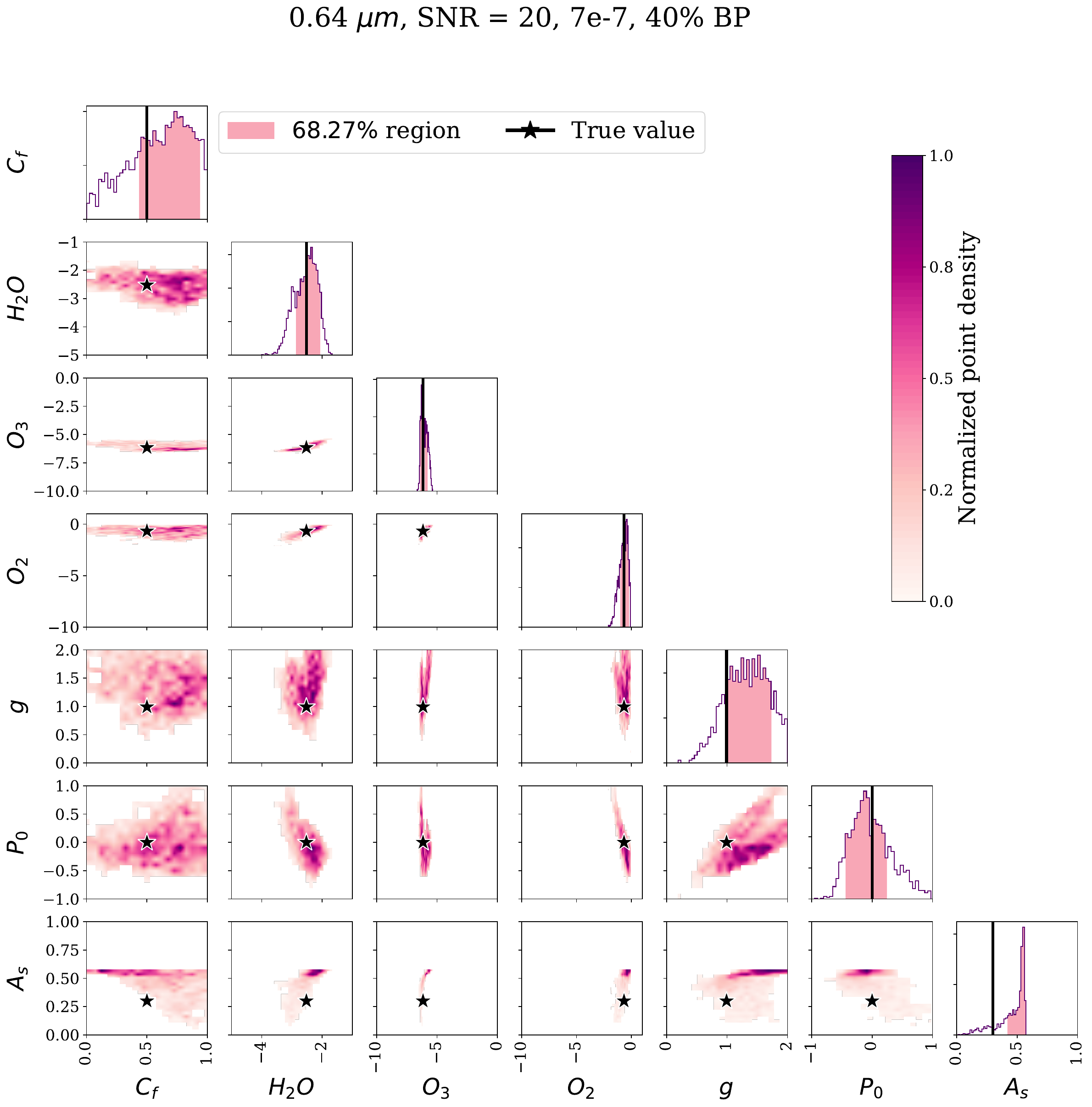}{0.5\textwidth}{(b) 40\% Bandpass, modern Earth values.}}
\gridline{\fig{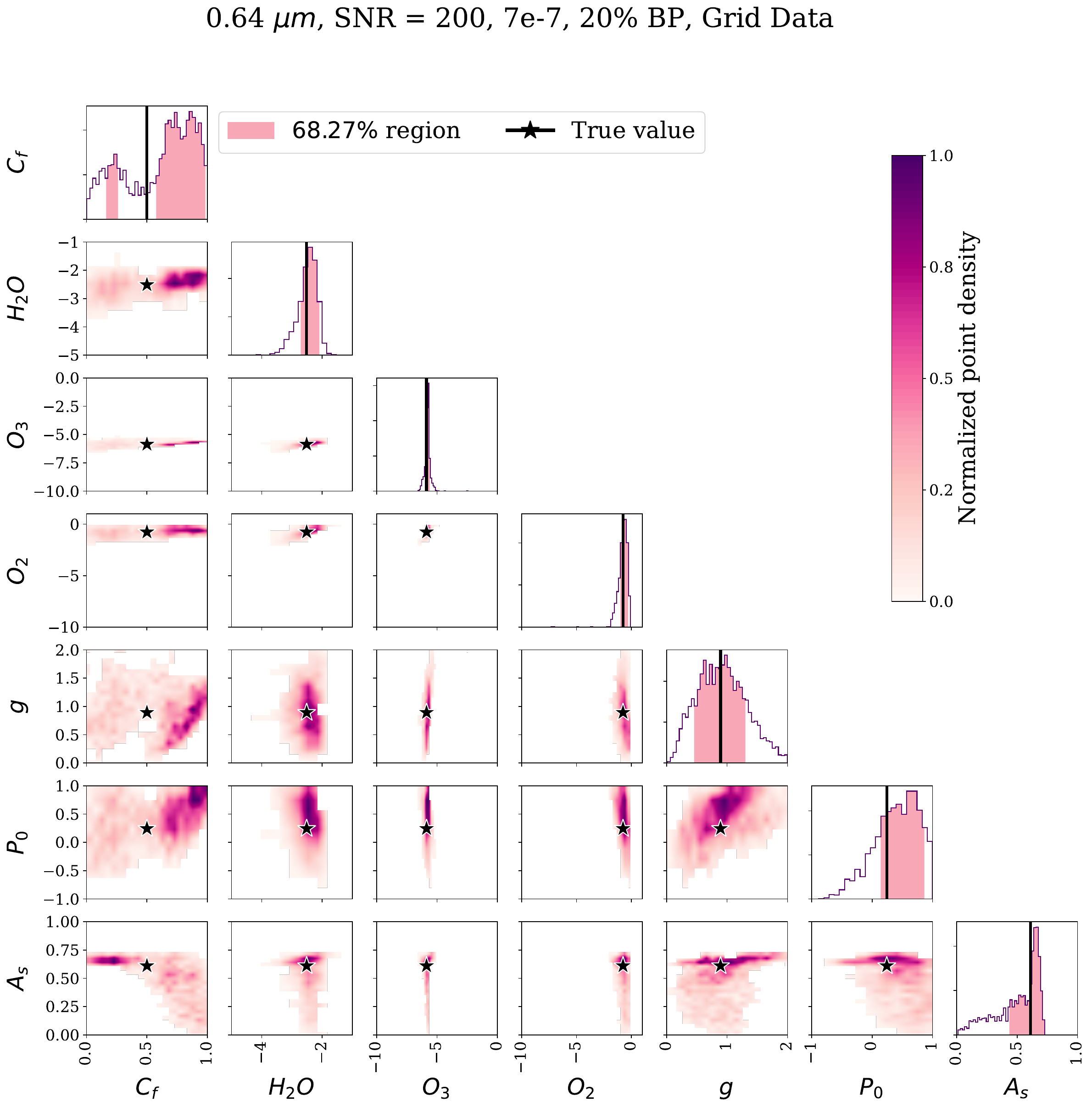}{0.5\textwidth}{(c) 20\% Bandpass, SNR = 200. Here we used only values that are directly from the grid itself to mitigate interpolation error, thus they are not modern Earth values.}
          \fig{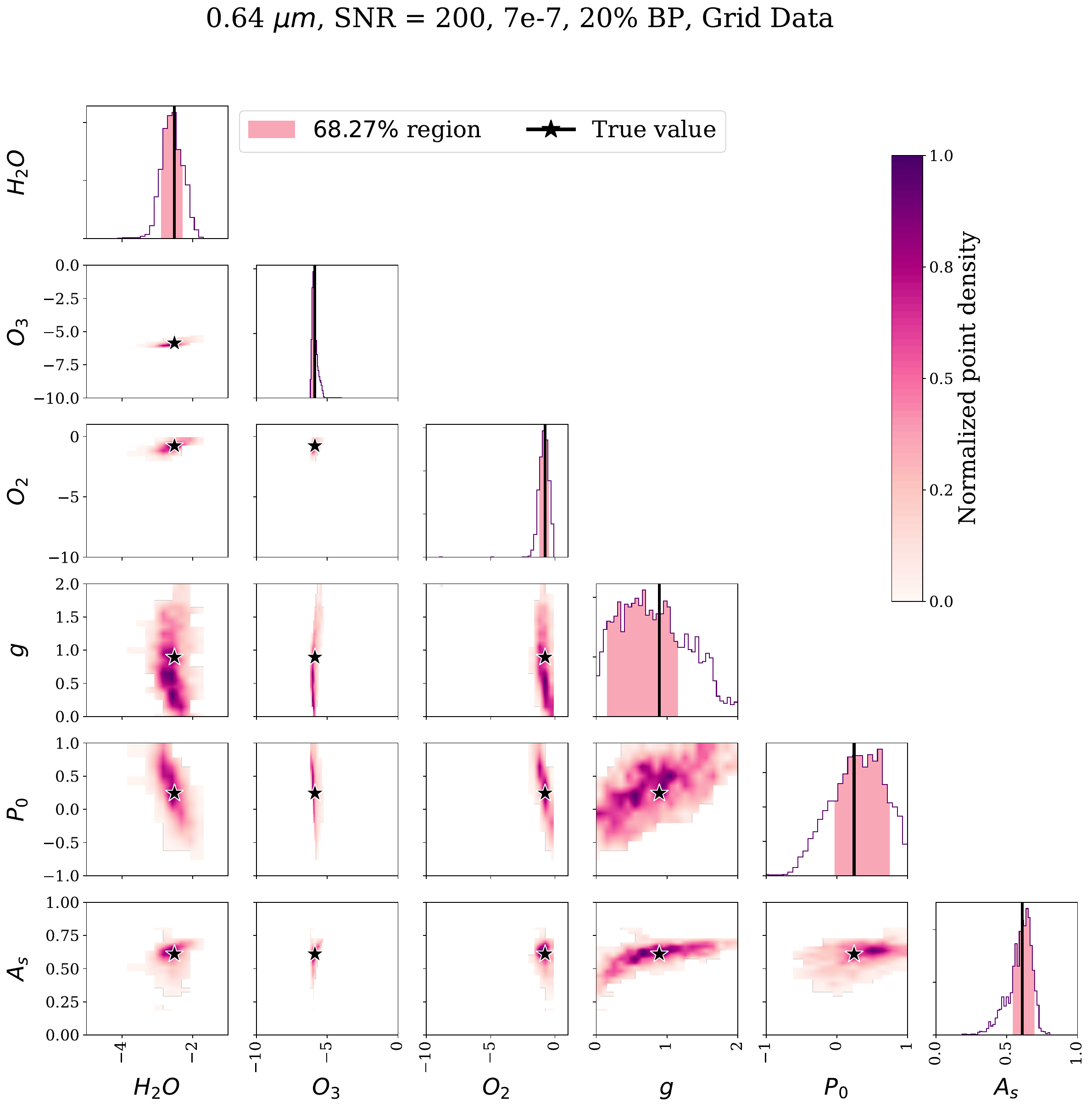}{0.5\textwidth}{(d) 
          20\% Bandpass, SNR = 200. Replicating \ref{fig:o3_corner}c, shown here are grid point values and not modern Earth values, with the notable difference that $C_{f}$ is locked to remove a known degeneracy source.}}
\centering
\caption{Corner plots for \ce{O3} at 0.64 {\microns} at SNR=20 to portray the lack of continuum and its impact on \ce{O3} abundance retrieval. We also show a corner plot for SNR = 200 with all parameters set to exact grid points to portray the interpolation error in $A_{s}$. The 68\% credible region is shown as pink shading in the 1D marginalized posterior distributions along the diagonal of the corner plot, and the true values are represented by black lines in the diagonals of the corner plot, and black stars within the 2D plots. It is clear that the error in A$_S$ is due to a combination of parameter degeneracy and interpolation error in the grid-derived forward models.}
\label{fig:o3_corner}
\end{figure*}

\section{Results}
\label{sec:results}

\subsection{Modern Earth Case Results}
\label{sec:modern}
We begin by presenting the detectability of \ce{O2} and \ce{O3} as a function of SNR for the fiducial modern-Earth case, as first examined by S23 and BARBIE1; all of the \ce{O2} and \ce{O3} data and calculated log-Bayes factors across abundance, SNR, and wavelength are available to the community on Zenodo\footnote{10.5281/zenodo.8349974}. We only present the narrow SNR range within which detectability strength changes materially for \ce{O2} and \ce{O3} in Figures~\ref{fig:o2_bfretest_narrow} and \ref{fig:o3_bfretest_narrow}. We can see that based on Figure~\ref{fig:o2_bfretest_narrow}, for SNR $\leq7$ there is only a weak or unconstrained detection possible for \ce{O2}. However, beginning at SNR = 7.5, we can see strong detections of \ce{O2} corresponding to bandpasses containing deep \ce{O2} spectral features, such as at 0.74 {\microns}. A strong detection of \ce{O2} becomes possible from 0.68 - 0.84 {\microns} by SNR = 8. At SNRs higher than 8, all bandpass locations yield a strong detection of \ce{O2} with increasingly better constraints of the 68\% credible region.

Looking to Figure~\ref{fig:o3_bfretest_narrow}, it was required to significantly increase the SNR to achieve a strong detection for \ce{O3}. We look at an SNR range of 18 -- 20, which is quite high, however it is only at this point that we begin to see a strong detection of \ce{O3} within the wavelength range of our simulations. At an SNR of 18, we can see a small area of weak detection between 0.6 and 0.67 {\microns} covering six bandpasses. At an SNR of 19, strong detection becomes possible for two bandpasses, and three bandpasses at SNR of 20. We can see it takes a very high SNR to achieve a strong detection of \ce{O3}, which would lead to an exceptionally high integration time. 

We can also see that although three bandpasses yield a strong detection at SNR=20, none of those bandpasses correctly constrain the abundance of \ce{O3} within the 68\% credible region. We present further investigation on this in Figure~\ref{fig:o3_corner}a. In this corner plot, which is focused on 0.64 {\microns} at a 20\% bandpass, \ce{O3} is not retrieved within the 68\% credible region, and there is a large spread of the high probability region which does not center on the true value of \ce{O3}. This is largely due to the lack of continuum caused by the depth and width of the ozone features; a similar problem for detecting \ce{H2O} at long wavelengths is discussed at length in BARBIE1. We present Figure~\ref{fig:o3_corner}b to show that by increasing the bandpass width from 20\% to 40\% centered on 0.64 {\microns}, and thus increasing the amount of spectral region covered in each bandpass, an adequate continuum is constrained and the retrieval of \ce{O3} is firmly within the 68\% credible region with little spread in the high probability regions. 

However, we can see in the corner plots shown in Figure~\ref{fig:o3_corner}a and \ref{fig:o3_corner}b that $A_{s}$ is also consistently retrieved at a value at least a factor of two away from the true value. To investigate the source of this incorrect retrieved parameter, we ran several retrievals with very high SNR (200) and with a data spectrum using values centered on a grid point, and compared both sets of results to our lower-SNR modern-Earth results.  This allowed us to test whether the result was due to degeneracies for lower-SNR data or due to the impact of interpolation error in our grid-based retrieval scheme (as examined by S23). In Figure~\ref{fig:o3_corner}c, where we use a very high SNR and all of the parameters are set to exact values found in the S23 grid, we can see that all parameters are retrieved within a 68\% credible region except for $C_{f}$ which is known to be degenerate. When $C_{f}$ is locked to its true value (0.5) as in Figure~\ref{fig:o3_corner}d, we can see this issue disappears and the range in the highest likelihood region decreases. There is higher interpolation error in $A_{s}$ likely due to the scarce sampling in the grid points and the limited bandwidth of the sampled spectra considered here. As there is only three grid points in this parameter space, there is a higher likelihood for interpolation error as there is more distance between the given true value and the nearest grid point. For a more in-depth description of this interpolation error and its impact, see S23. 

\begin{figure*}
\centering
\gridline{\fig{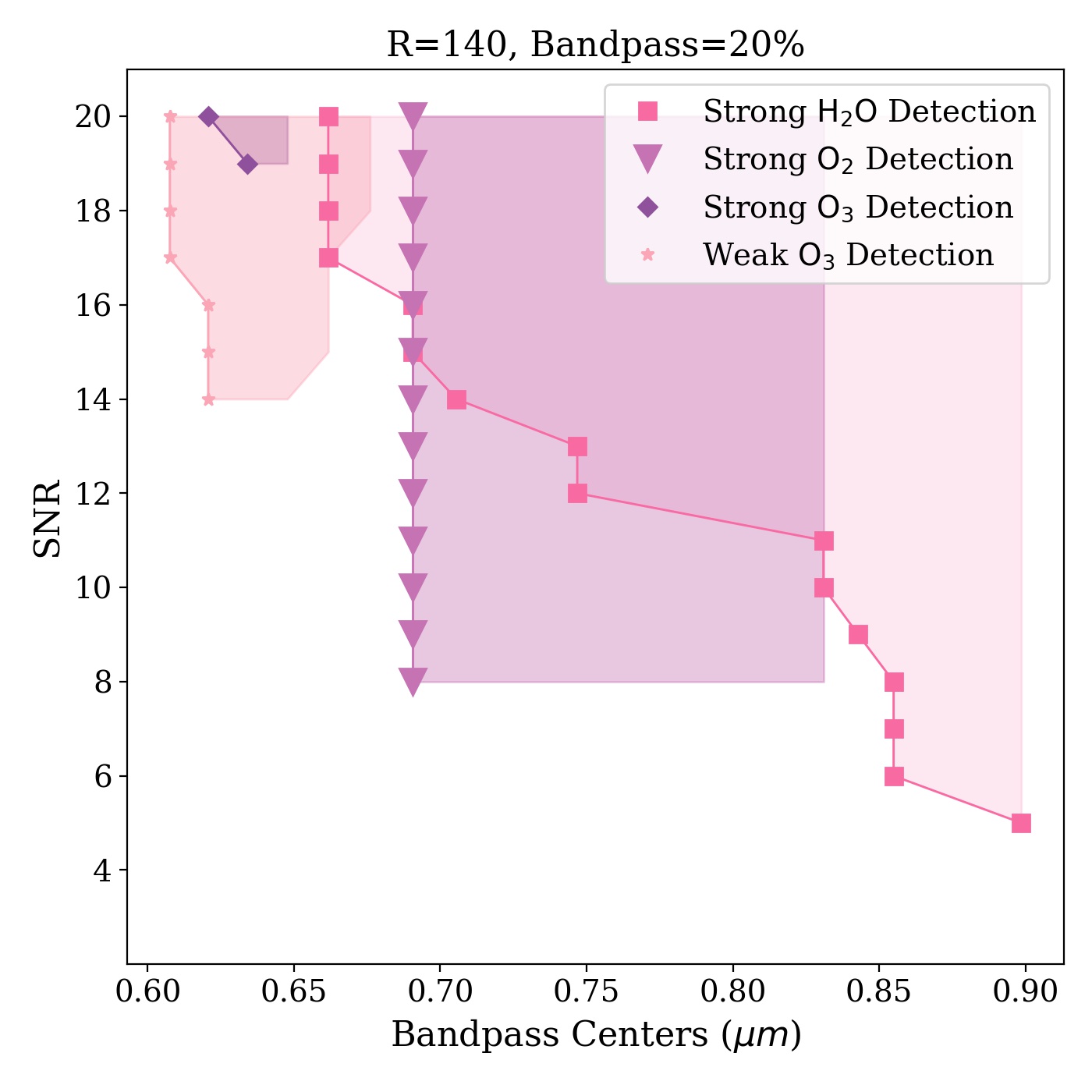}{0.334\textwidth}{(a)      
          Summary for Figures~\ref{fig:o2_bfretest_narrow} and \ref{fig:o3_bfretest_narrow} with 20\% bandpasses.}
          \fig{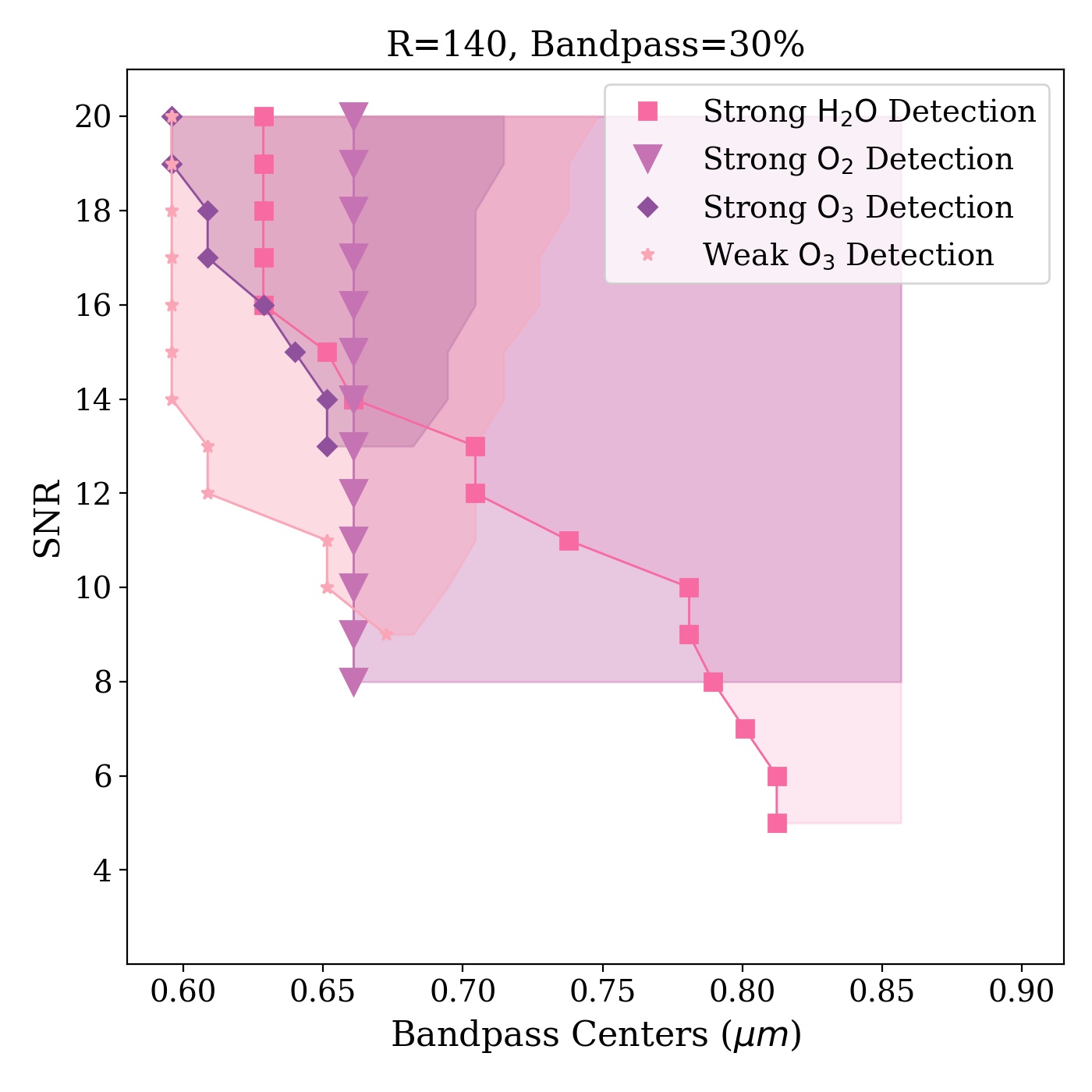}{0.334\textwidth}{(b) Same, but with 30\% bandpasses.}
          \fig{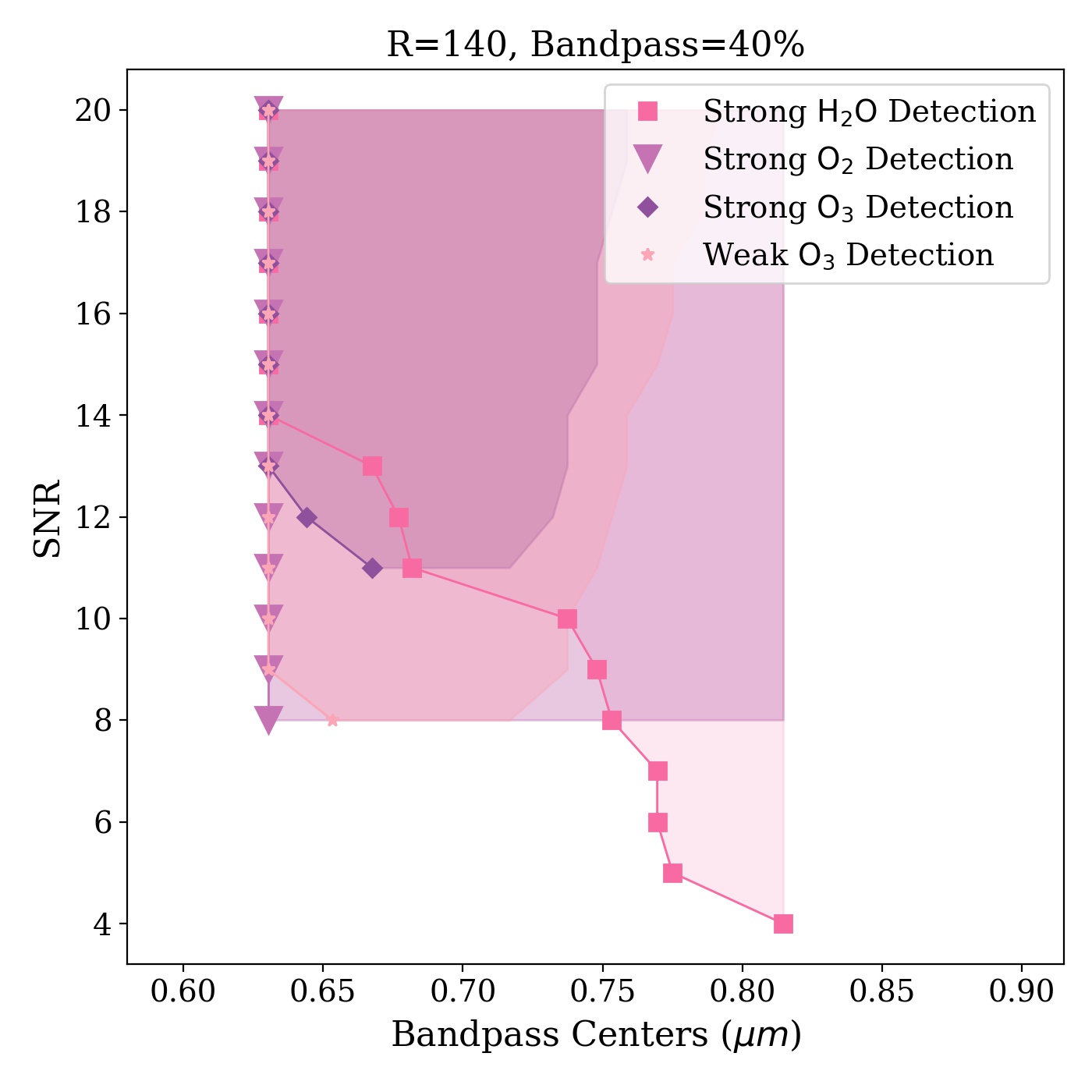}{0.334\textwidth}{(c) Same, but with 40\% bandpasses.}}
\caption{In all panels, the shortest bandpass center at which one can achieve a strong detection for \ce{O2}, or a strong or weak detection for \ce{O3} are shown in purple triangles, dark purple diamonds, and light pink dots respectively. The strong detection range for \ce{H2O} is presented in pink squares, to provide context for dual or triple molecule detection. The range between the shortest and longest bandpass center at which detection is possible is filled in. We present the range for a strong detection SNR is on the y-axis, and the bandpass centers are on the x-axis. Note that the last bandpass center for strong detection is shaded out to the long edge of the bandpass, to maintain ability to directly compare the panels.}
\label{fig:shortest_detec}
\end{figure*}

In Figure~\ref{fig:shortest_detec} we present the shortest wavelength at which a weak or strong detection is achieved for \ce{O2} and \ce{O3}; as described in BARBIE1, this is an important metric since the number of planets amenable to high-contrast coronagraphic imaging is higher when observing at shorter wavelengths due to the smaller inner working angle. Figure~\ref{fig:shortest_detec}a provides a summarial result of Figures~\ref{fig:o2_bfretest_narrow} and \ref{fig:o3_bfretest_narrow}, covering the full range of SNR from 3-20 at a 20\% bandpass. In Figure~\ref{fig:shortest_detec}b, we present the shortest wavelength for strong detection if we assume a 30\% bandpass, while in Figure~\ref{fig:shortest_detec}c we present the shortest wavelength for strong detection if we assume a 40\% bandpass. We also present the range for a strong detection of \ce{H2O} as in BARBIE1, with additional SNRs to 20, to provide context to the results and present the possibilities for dual or triple molecule detection. To show the full range, we shade out to the longest wavelength where detection is possible. 

In Figure~\ref{fig:shortest_detec}a we can see that \ce{O3} is only detectable, whether weakly or strongly, with high SNR ($\ge14$) data over a very narrow range. We can see that \ce{O3} can be detected (albeit weakly) simultaneously with a strong detection of \ce{H2O} at high SNR ($\ge17$) with careful bandpass selection. However, the high SNR required for detection indicates that this would be costly in terms of observing time. Conversely, \ce{O2} is strongly detectable at an SNR of 8, with the range between shortest and longest wavelength encompassing all \ce{O2} features in the visible wavelength range. This detectability range overlaps significantly with the strong detection range of \ce{H2O}. We can see that \ce{O2} and \ce{H2O} can be simultaneously observed with SNR = 10 with careful selection of wavelength and corresponding bandpass. This also allows for a range of possible SNR depending on the desired wavelength of detection - if a longer wavelength such as 0.83 {\microns} is accessible, then SNR = 10 would allow for a dual detection, but if a short wavelength such as 0.68 {\microns} is required, then a much higher SNR is required. Next looking to Figure~\ref{fig:shortest_detec}b, we see the detectability change, with triple molecule detection possible from a shortest wavelength of 0.66 {\microns} out to 0.7 {\microns} at an SNR $\ge13$. \ce{O2} remains strongly detectable beginning at an SNR of 8 as in Figure~\ref{fig:shortest_detec}a; however, while the shortest wavelength of detectability for \ce{O2} starts at approximately 0.69 {\microns} with a bandpass of 20\%, the shortest wavelength of detectability for \ce{O2} with a bandpass of 30\% starts at approximately 0.66 {\microns}. When we look to Figure~\ref{fig:shortest_detec}c, we see that detectability changes drastically, with triple molecule strong detection possible from a shortest wavelength of 0.625 {\microns} out to 0.725 {\microns} at an SNR $\ge11$. We can also see that once again, \ce{O2} remains strongly detectable beginning at an SNR of 8 as in Figure~\ref{fig:shortest_detec}a and \ref{fig:shortest_detec}b, the shortest wavelength of detectability for \ce{O2} changes once more, starting at approximately 0.625 {\microns} with a bandpass of 40\%. Thus, although the SNR of strong detectability does not change, the minimum possible wavelength of strong detectability significantly decreases as a function of bandpass width. 

\subsection{Results for Varying Abundance Cases}
\label{sec:abun}

\begin{figure*}
\centering
\gridline{\fig{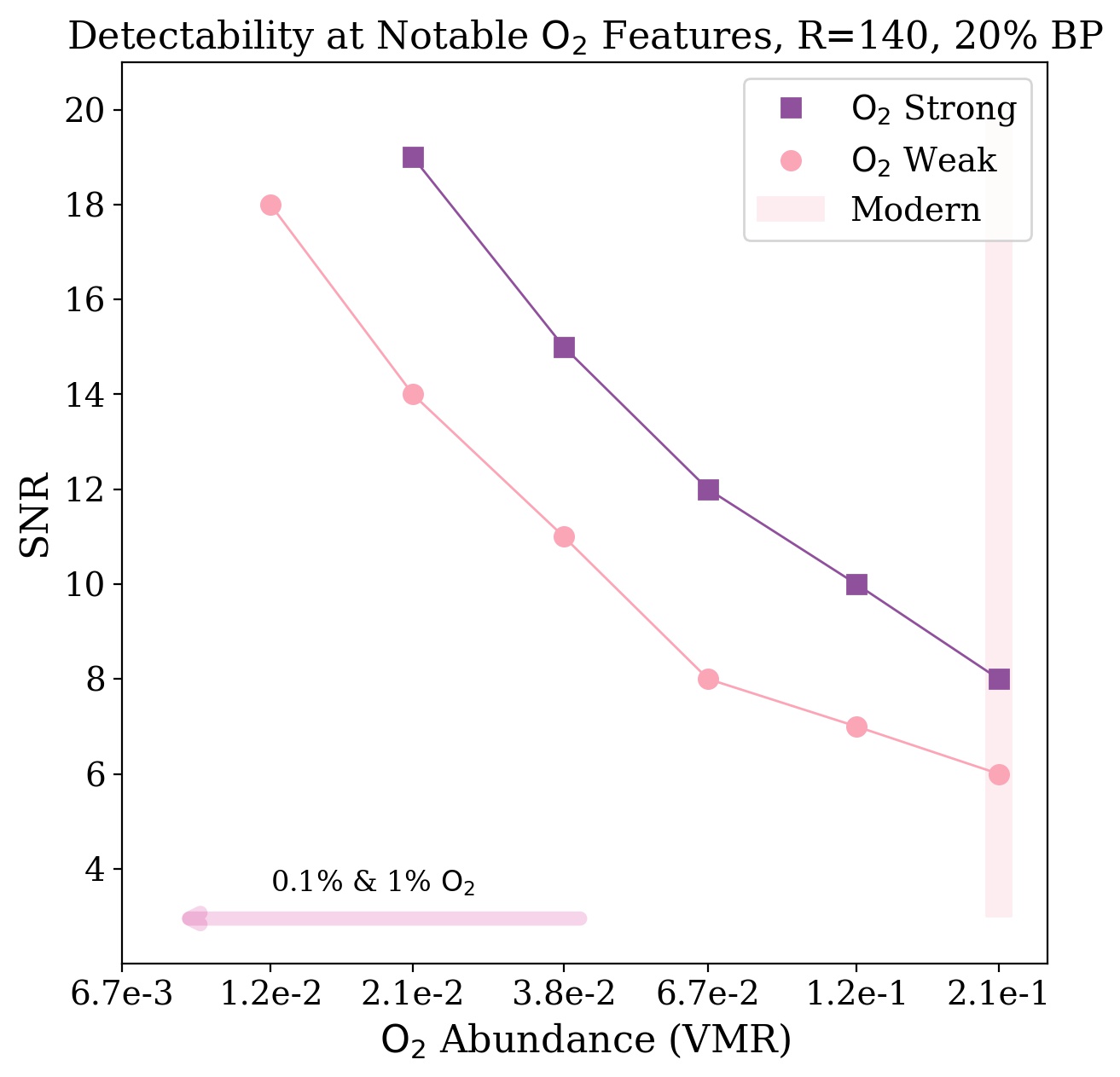}{0.334\textwidth}{(a)     
          Shown above is the detectability of the 0.74 {\microns} \ce{O2} feature spanning multiple abundances and SNRs with a 20\% bandpass.}
          \fig{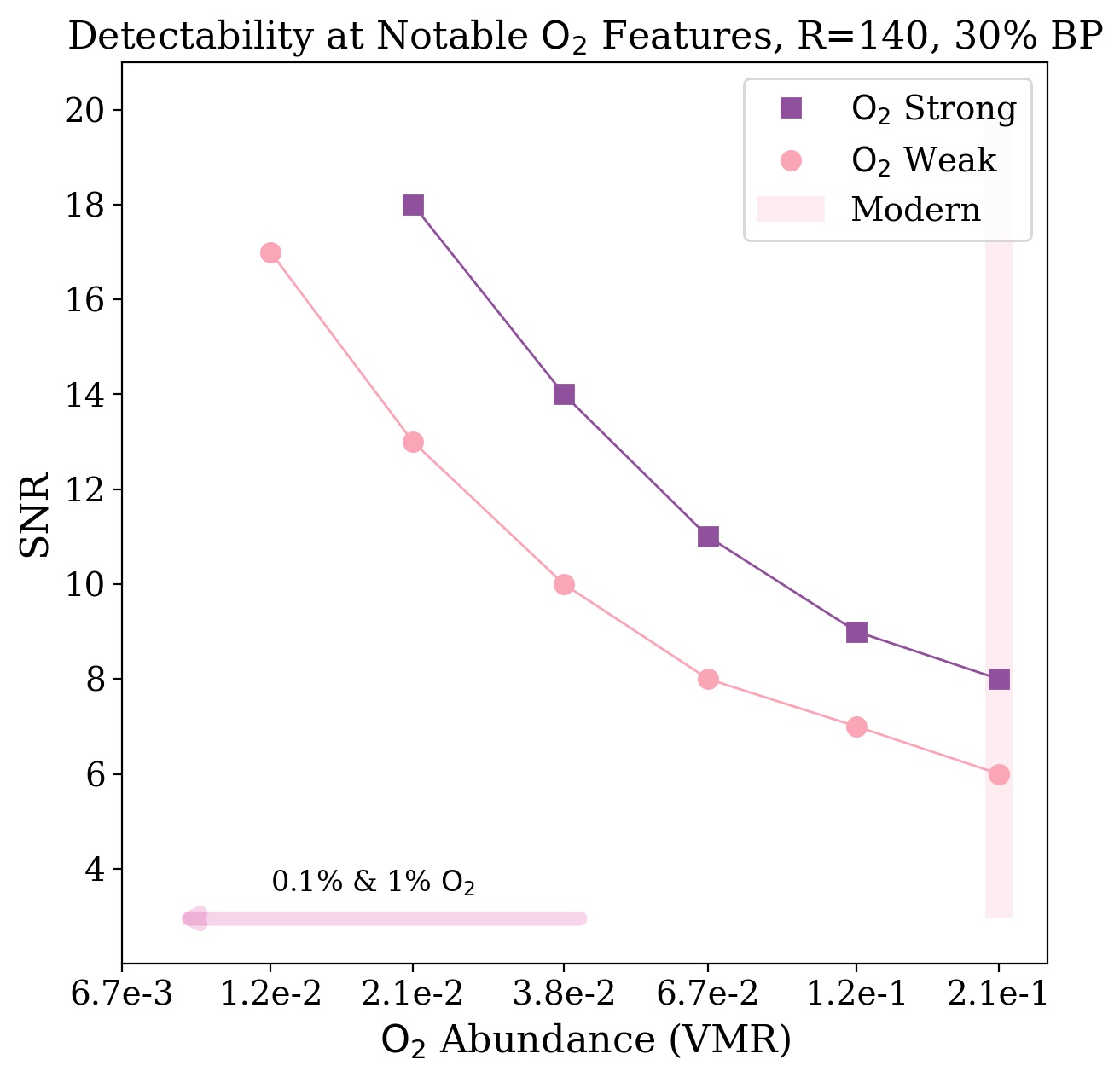}{0.334\textwidth}{(b) Same, but a 30\% bandpass.}
          \fig{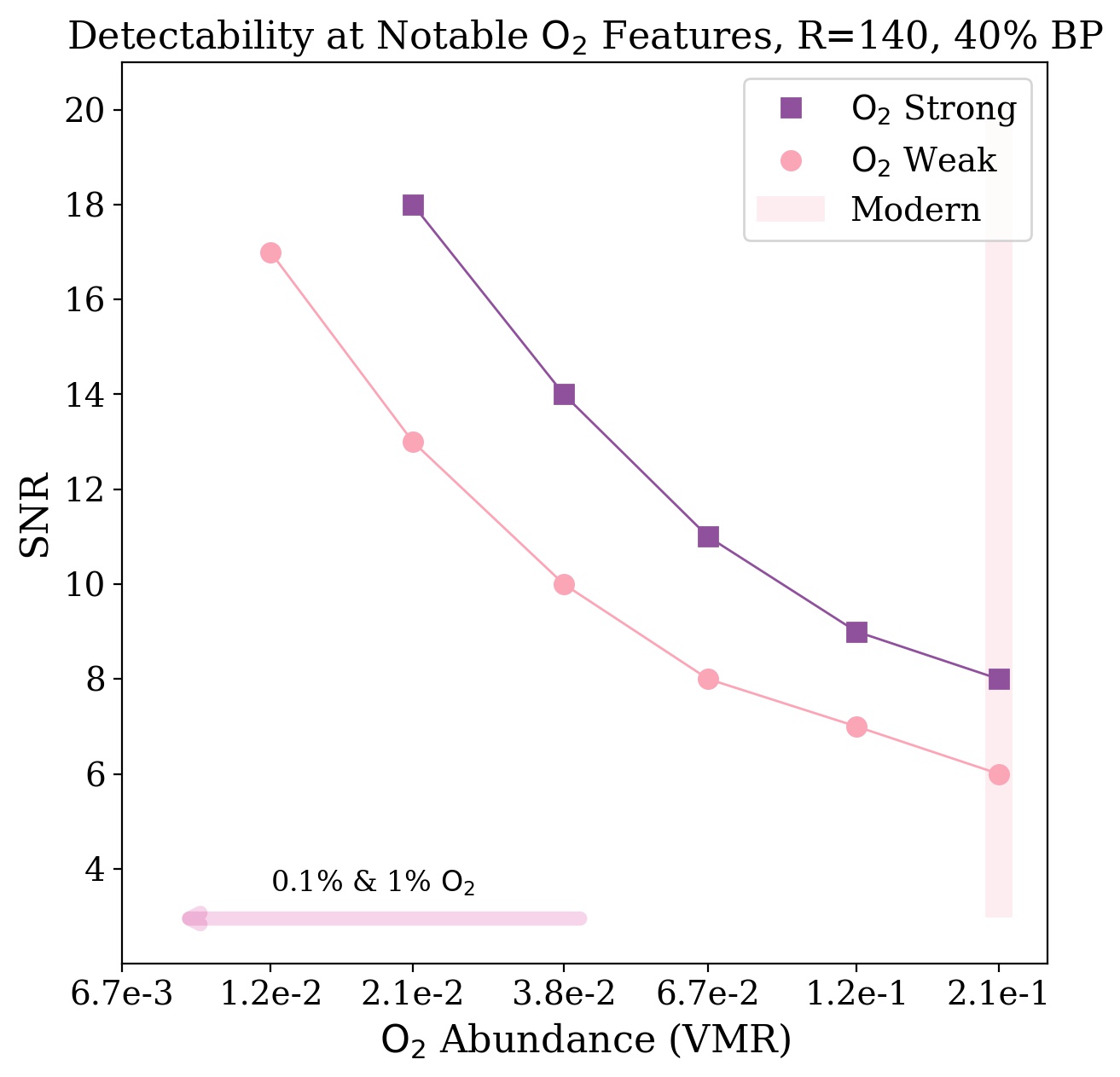}{0.334\textwidth}{(b) Same, but a 40\% bandpass.}}
\caption{Shown above is the lowest SNR values for strong detection of the 0.74 {\microns} \ce{O2} feature as a function of \ce{O2} abundance. We note that although the 0.1\% and 1\%  modern \ce{O2} abundances (i.e. mid-Proterozoic abundances) are in our study, they are not presented on the plot. We denote these abundances with a pink arrow. The VMR values are on the x-axis, with SNR on the y-axis. Strong detection is shown in purple squares, weak detection is shown in pink dots, the modern abundance range is highlighted with a light pink strip.}
\label{fig:epoch_o2}
\end{figure*}

At this point in our study, we shift to present our abundance case study, wherein we vary the abundance of \ce{O2} below modern-Earth values and \ce{O3} above modern-Earth values. To assess the trade-off between longer observations (i.e., higher SNR) and different concentrations of \ce{O2} and \ce{O3}, we varied the SNR on the observations for the full range of different VMRs per molecule. In Figure~\ref{fig:epoch_o2}a, \ref{fig:epoch_o2}b, and ~\ref{fig:epoch_o2}c, we display the minimum SNR required to achieve a strong detection for each abundance of \ce{O2} at 0.76 {\microns} at 20\%, 30\%, and 40\% bandpasses respectively. Looking first to Figure~\ref{fig:epoch_o2}a, we notice that \ce{O2} quickly requires mid- to high-SNR data to be strongly or weakly detected with even one order of magnitude decrease in abundance. In fact, many of the abundances in our simulation are not detectable at all in this SNR range. The Proterozoic abundances of \ce{O2} (0.1\% to 1\% of modern Earth abundance) are not detectable at any SNR $\le20$, and in fact will likely require an extremely high SNR to become detectable, as these values are two to three magnitudes lower than modern Earth abundance values. As seen in Figure~\ref{fig:epoch_o2}b, varying the bandpass to 30\% decreases the required SNR for detection across almost all abundances of \ce{O2}. For instance, where at a 20\% bandpass it requires an SNR of 19 to strongly detect \ce{O2} at 2.1$\times10^{-2}$ VMR, with a 30\% bandpass the required SNR drops to 18. This is true for all detectable abundance cases except modern Earth values, which consistently requires an SNR of 8 for strong detection. In Figure~\ref{fig:epoch_o2}c, we do not see a difference in required SNR for strong detection with a 40\% bandpass.  

\begin{figure*}
\centering
\includegraphics[scale=0.5]{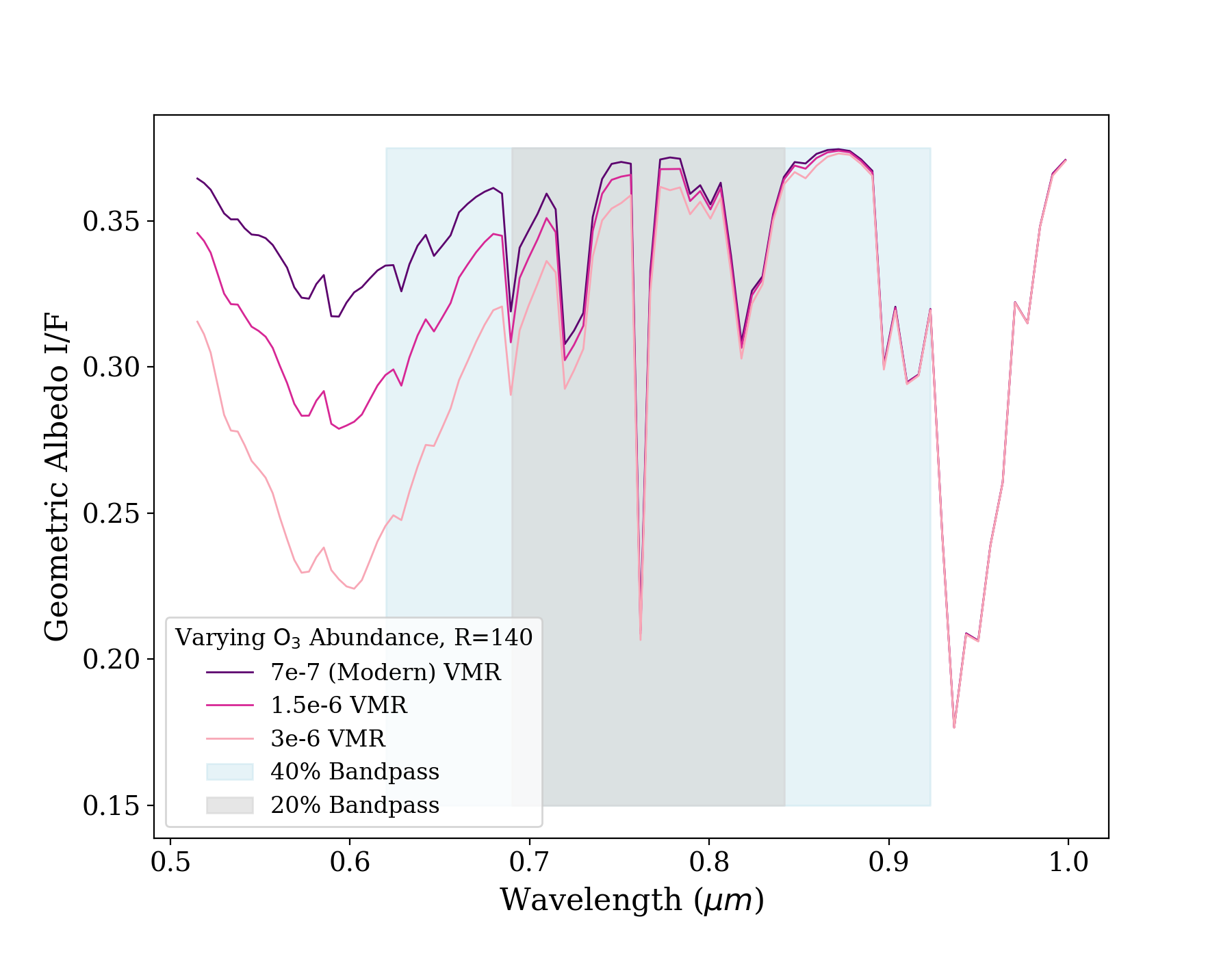}
\caption{Spectra with varied \ce{O3} abundances. We feature the highest abundance (3$\times10^{-6}$), the mid point abundance (1.5$\times10^{-6}$), and lowest abundance in the study (7$\times10^{-7}$) (modern Earth), respectively. All spectra are binned at R=140. We also present a 20\% and 40\% bandpass range width, in light grey and light blue respectively.}
\label{fig:o3_abun}
\end{figure*}

\begin{figure*}
\centering
\gridline{\fig{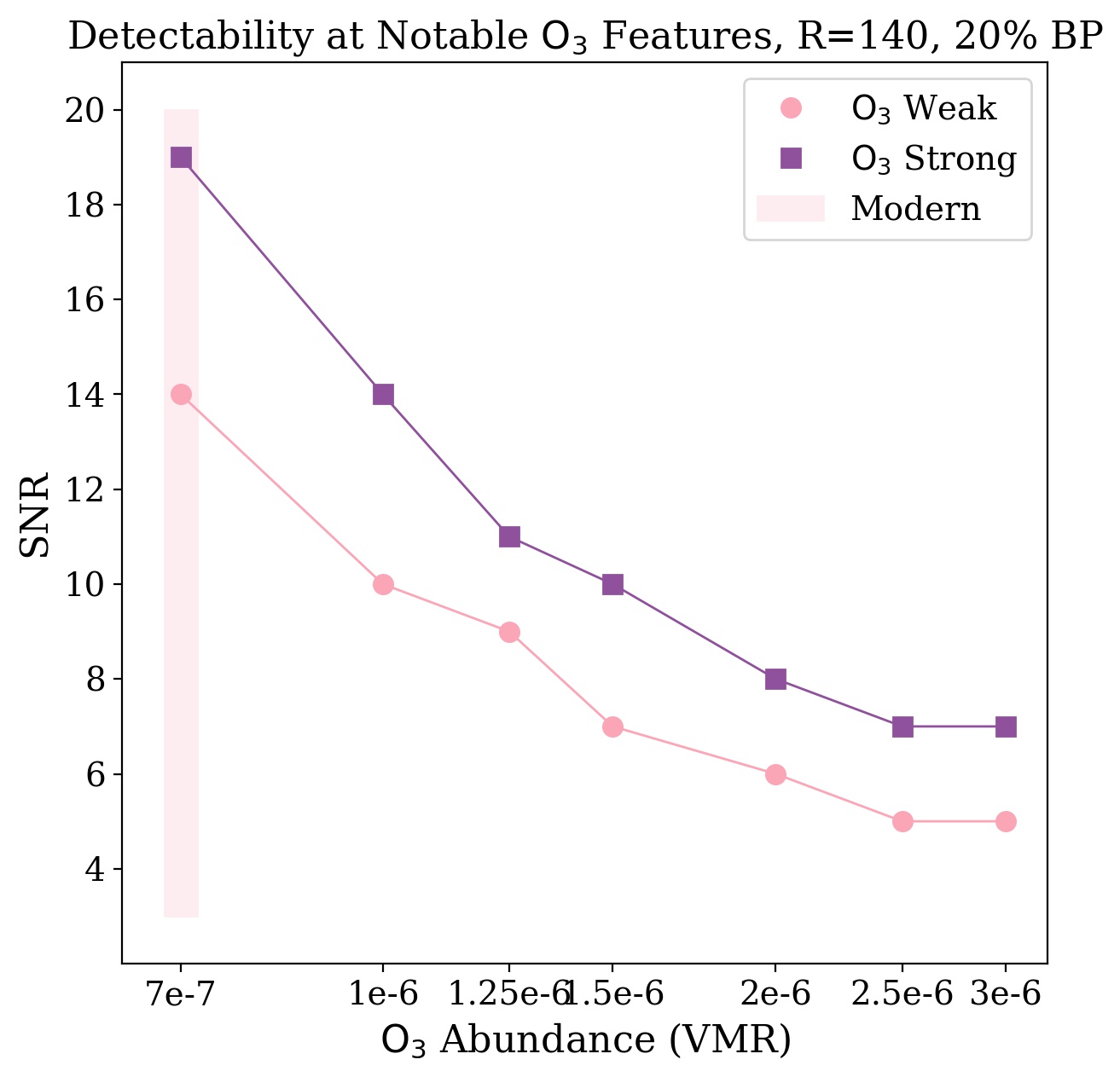}{0.334\textwidth}{(a)     
          Shown above is the detectability of the 0.63 {\microns} \ce{O3} feature spanning multiple abundances and SNRs with a 20\% bandpass.}
          \fig{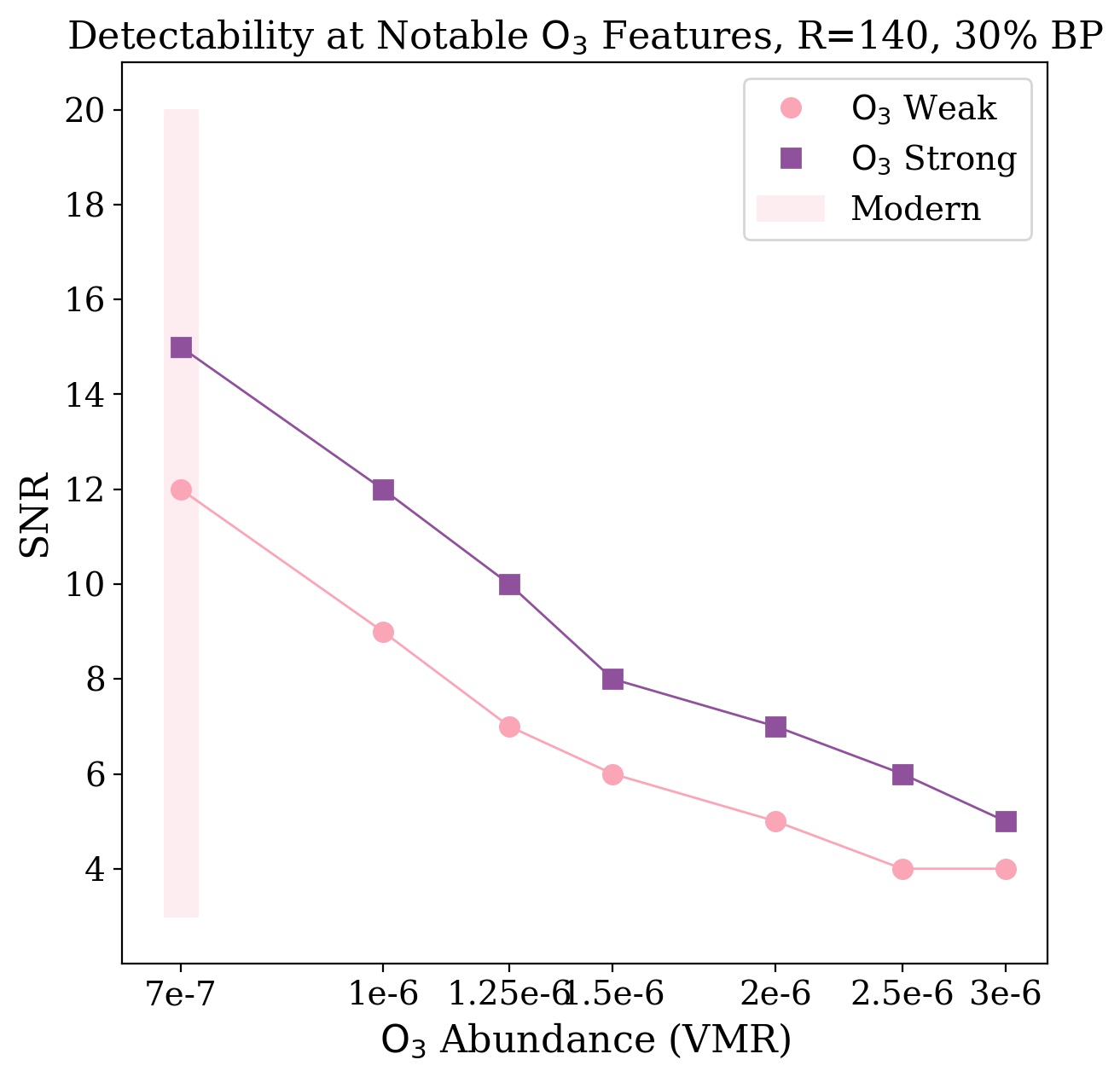}{0.334\textwidth}{(b) Same, but a 30\% bandpass.}
          \fig{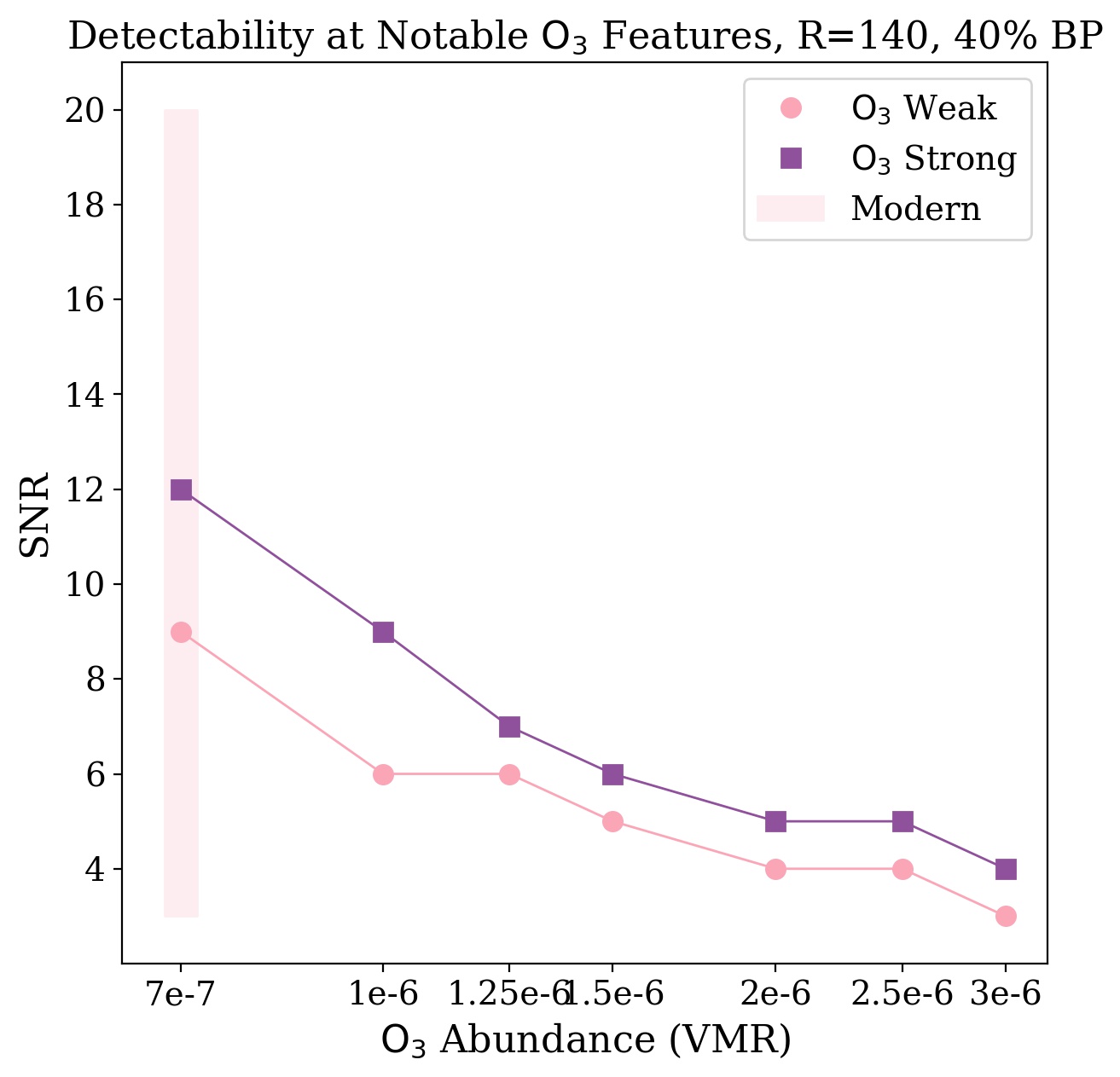}{0.334\textwidth}{(b) Same, but a 40\% bandpass.}}
\caption{Shown above is the lowest SNR values for strong detection of the 0.63 {\microns} \ce{O3} feature as a function of \ce{O3} abundance. The VMR values are on the x-axis, with SNR on the y-axis. Strong detection is shown in purple squares, weak detection is shown in pink dots, and the modern abundance range is highlighted with a light pink strip.}
\label{fig:epoch_o3}
\end{figure*}

While bandpass width makes little difference to the detectability of \ce{O2}, it makes a significant difference when detecting \ce{O3}. We can see in Figure~\ref{fig:o3_abun} the large impact of the change in bandpass width. With a 20\% bandpass centered on the 0.76 {\microns} \ce{O2}, we capture the entirety of the \ce{O2} feature, along with the two smaller \ce{H2O} features at 0.74 and 0.84 {\microns}. When the bandpass is widened to 40\%, we capture all off the above, along with a portion of both the \ce{O3} feature that peaks at approximately 0.63 {\microns} and the 0.9 {\microns} \ce{H2O} feature. In Figure~\ref{fig:epoch_o3}a, \ref{fig:epoch_o3}b, and \ref{fig:epoch_o3}c, we display the minimum SNR required to achieve a strong detection for each abundance of \ce{O3} at 0.64 {\microns} with 20\%, 30\%, and 40\% bandpasses respectively. In Figure~\ref{fig:epoch_o3}a, we can see that \ce{O3} is detectable, strongly and weakly, in the full range of our simulation values. At modern Earth abundances, the required SNR for a strong detection is high at 19, but it is possible to achieve a weak detection at SNR = 14. When we increase to high \ce{O3} values, approximately 5x higher in abundance than modern Earth, the required SNR drops drastically, with strong or weak detection requiring SNRs of 7 or 5 respectively. The largest decrease in required SNR occurs between modern Earth values (7$\times10^{-7}$ VMR) and 1.25$\times10^{-6}$ VMR, dropping steeply from a required SNR of 19 to 11 for a strong detection, and 14 to 9 for a weak detection. When we increase the bandpass width to 30\% as in Figure~\ref{fig:epoch_o3}b, we see the required SNR for detection drop across all abundance values. For instance, at 1.5$\times10^{-6}$ VMR, the required SNR for strong detection at a 20\% bandpass is 10, with a 30\% bandpass the required SNR drops to 8. This is true for all values across the abundances cases, to varying degrees of severity. This happens once more when the bandpass is widened to 40\% as in Figure~\ref{fig:epoch_o3}c. Looking to the same example case of 1.5$\times10^{-6}$ VMR, the required SNR for strong detection drops to 6 at a 40\% bandpass. With each increase of bandpass width, strong detectability at mid-low SNR data becomes more accessible, with a required SNR as low as 4 for a strong detection at 3$\times10^{-6}$ VMR.  

\subsection{Self-Consistent Chemistry Comparison}
\label{sec:kozakis}

\begin{table}[h]
    \centering
    \begin{tabular}{cccc} 
        \hline
        \hline
        \textbf{Parameter Symbol} & \textbf{Fiducial} & \textbf{New Values}\\
        \hline
        $\mathrm{C_f}$ & 0.5 \\
        \ce{H2O} & 3$\times10^{-3}$ VMR \\
        \ce{O3} & 7$\times10^{-7}$ VMR & 105\%, 110\%, 105\% PAL\\
        \ce{O2} & 0.21 VMR & 10\%, 50\%, 75\% PAL\\
        $\mathrm{P_0}$ & 1.0 Bar \\
        $\mathrm{g}$ & 9.8 m/s \\
        $\mathrm{A_s}$ & 0.3 \\
        \hline
    \end{tabular}
    \caption{New input data spectrum values for the chemically consistent retrievals, following the {\tt Atmos} values found in \citet{kozakis22} for \ce{O2} and \ce{O3}}
    \label{tab:atmosvals}
\end{table}

\begin{figure*}
\centering
\gridline{\fig{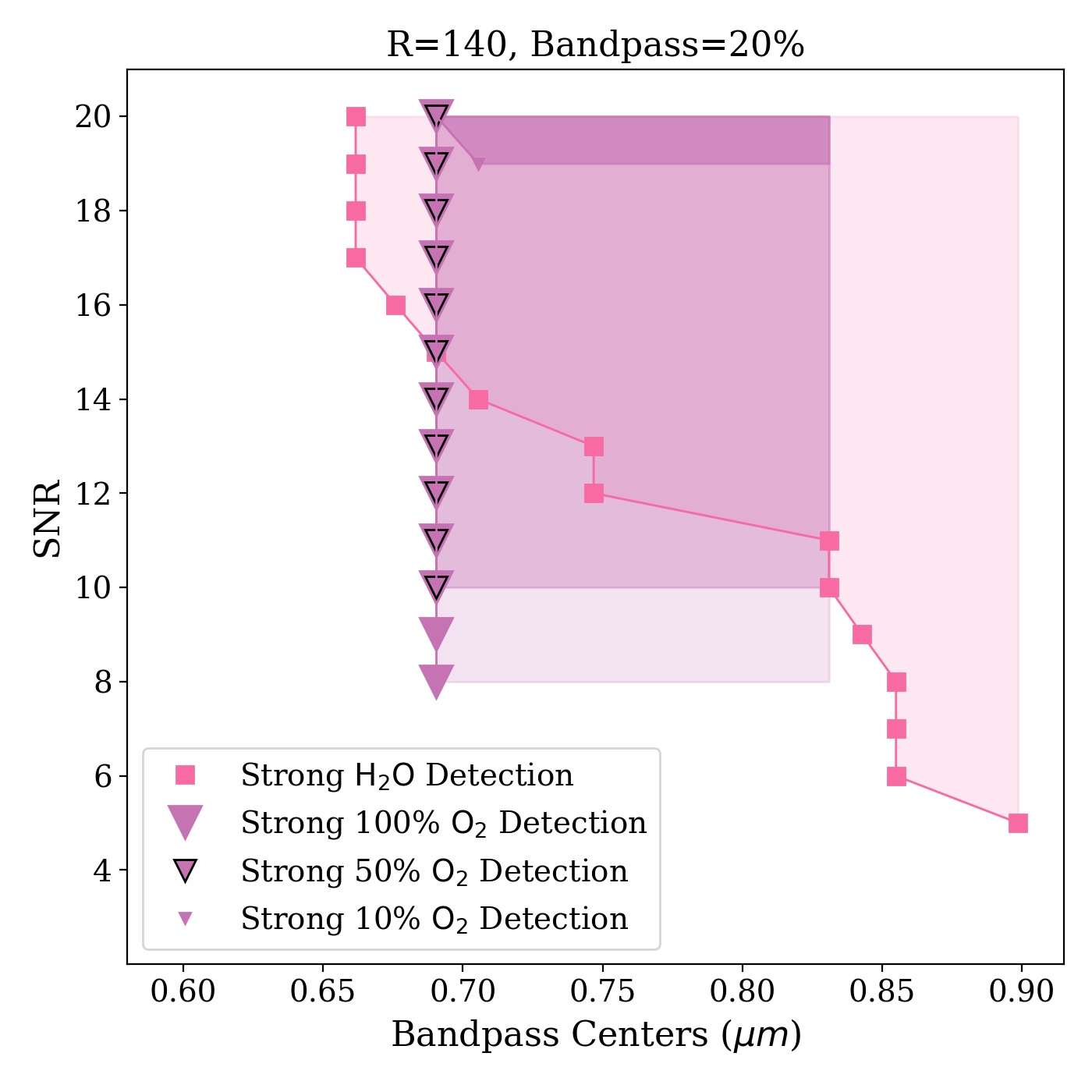}{0.334\textwidth}{(a)   
          \ce{O2} and \ce{H2O}, 20\% bandpass.}
          \fig{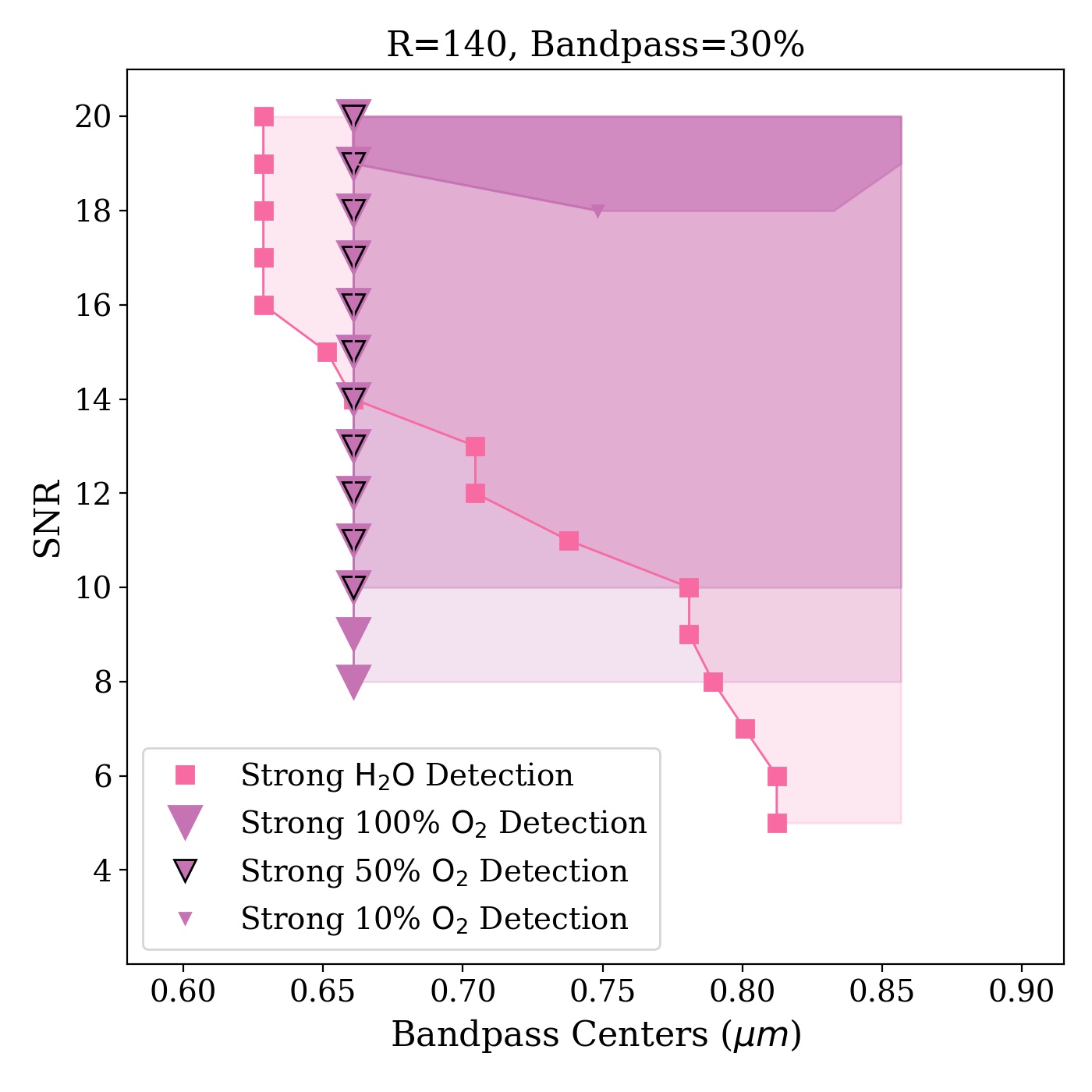}{0.334\textwidth}{(b) \ce{O2} and \ce{H2O}, 30\% bandpass.}
          \fig{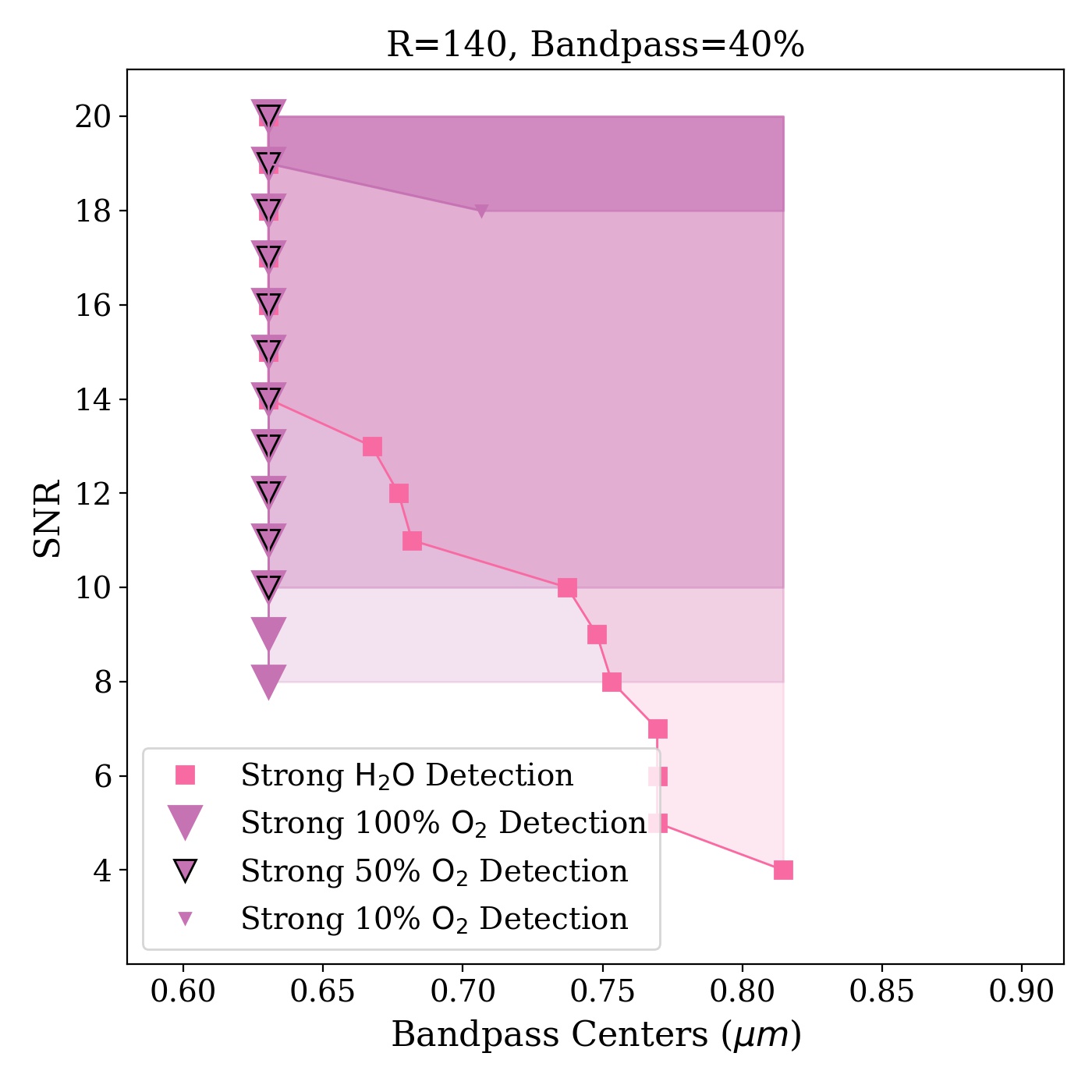}{0.334\textwidth}{(c) \ce{O2} and \ce{H2O}, 40\% bandpass.}}
\gridline{\fig{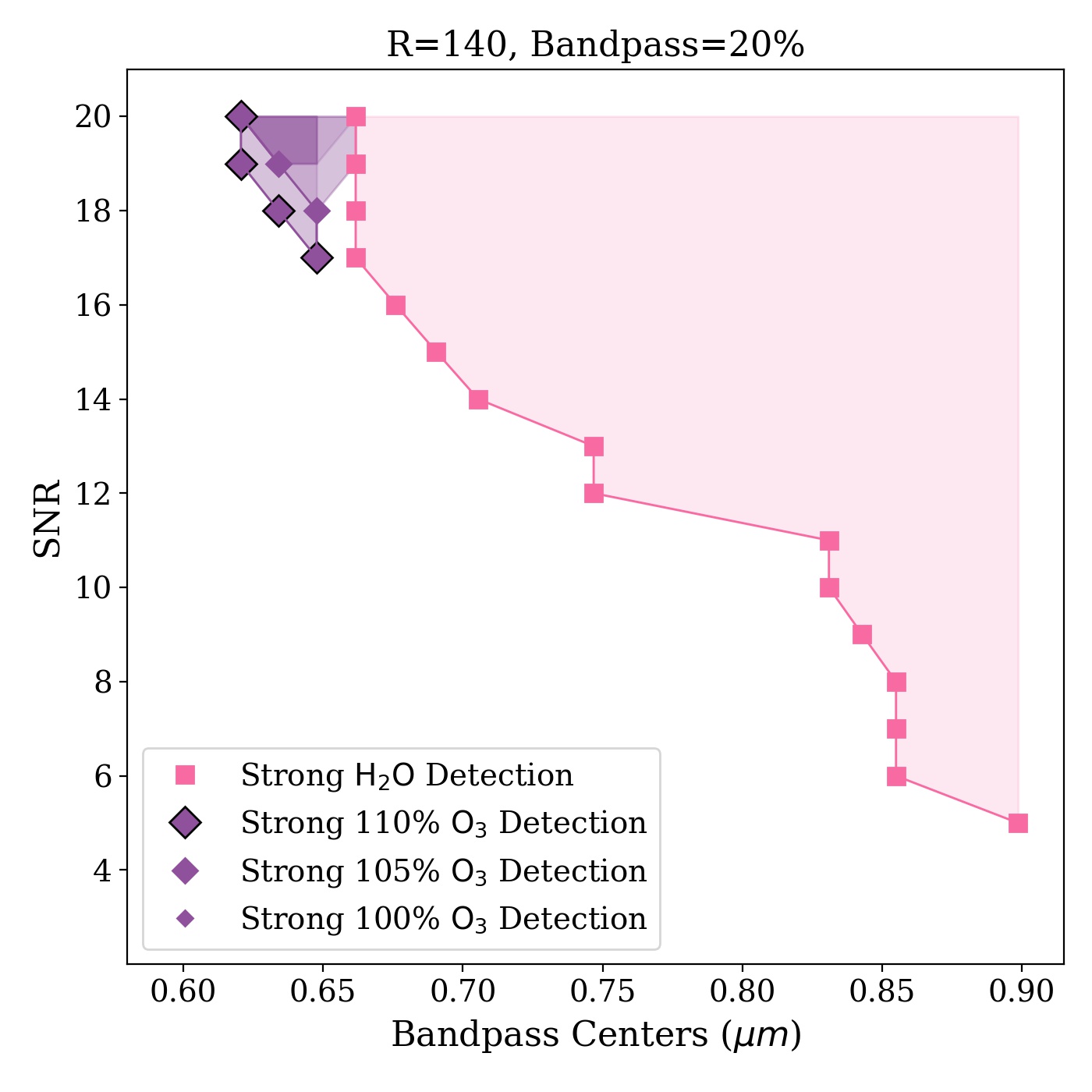}{0.334\textwidth}{(d)   
          \ce{O3} and \ce{H2O}, 20\% bandpass.}
          \fig{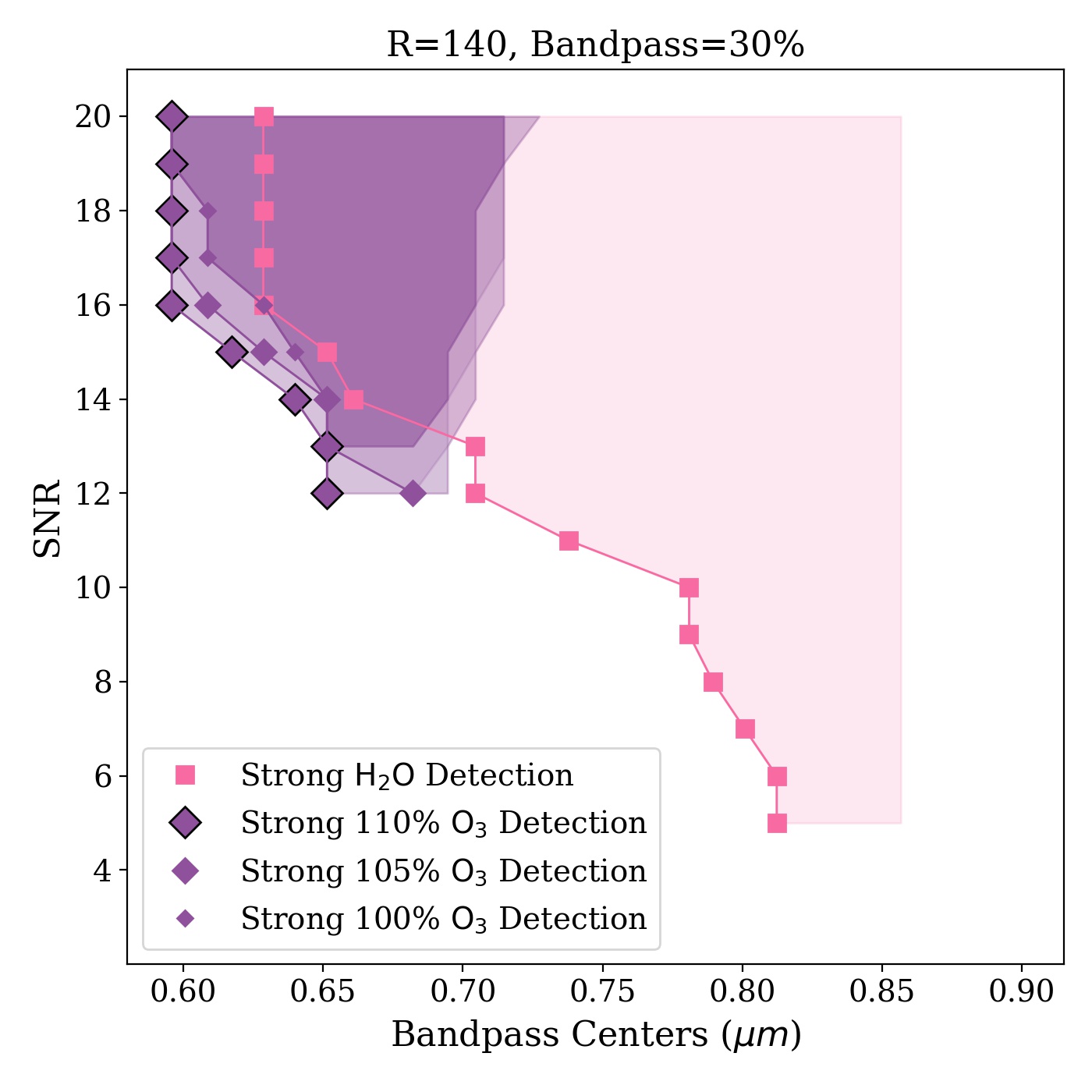}{0.334\textwidth}{(e) \ce{O3} and \ce{H2O}, 30\% bandpass.}
          \fig{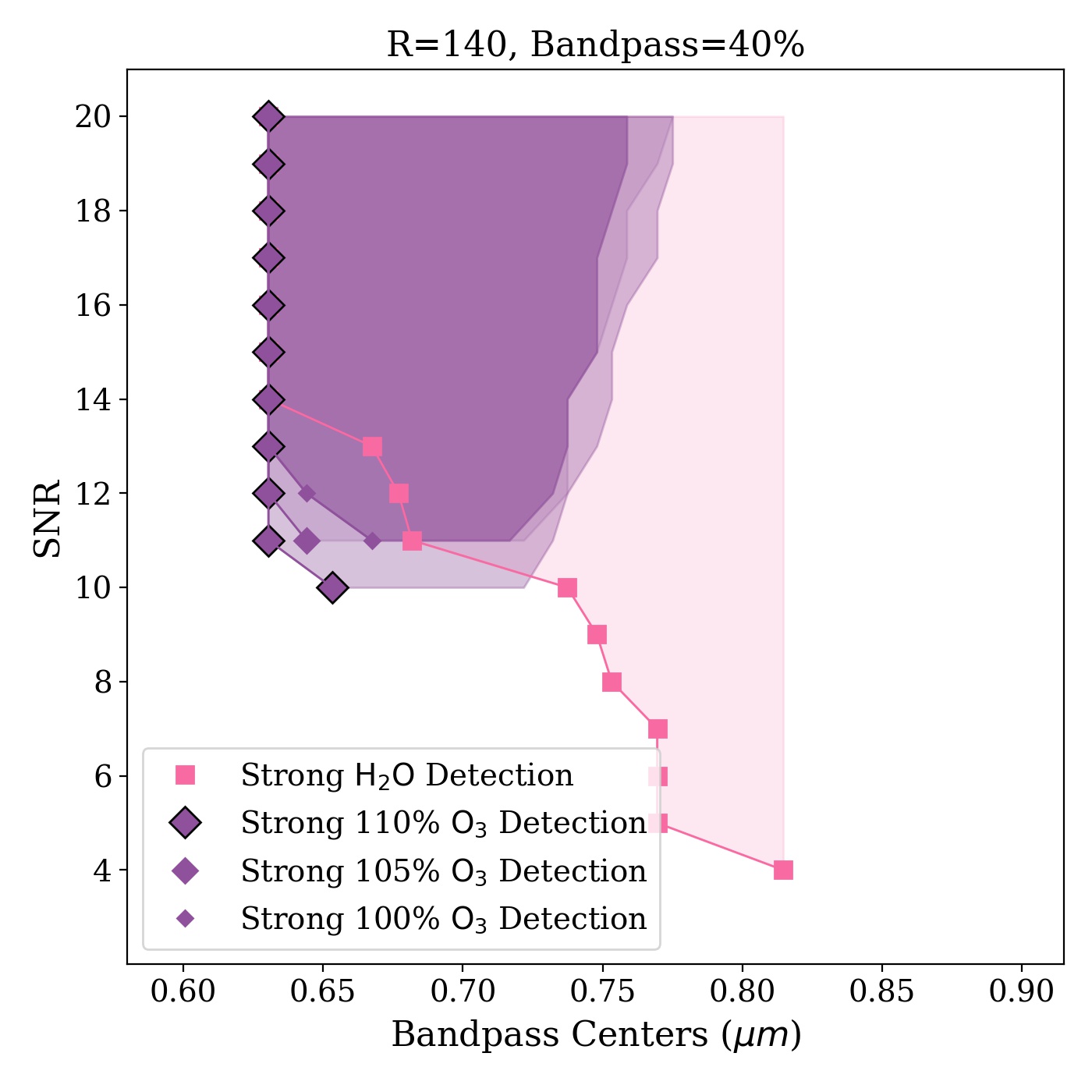}{0.334\textwidth}{(f) \ce{O3} and \ce{H2O}, 40\% bandpass.}}
\caption{The shortest bandpass center at which one can achieve a strong detection for \ce{O2} or \ce{O3} at varying present atmospheric levels (PALs) for a specific SNR are shown in purple triangles and dark purple diamonds respectively. The range between the shortest and longest bandpass center at which detection is possible is filled in. The strong detection range for \ce{H2O} is presented in pink squared to provide context to the results.}
\label{fig:o2o3testcomp}
\end{figure*}

\begin{figure*}
\centering
\gridline{\fig{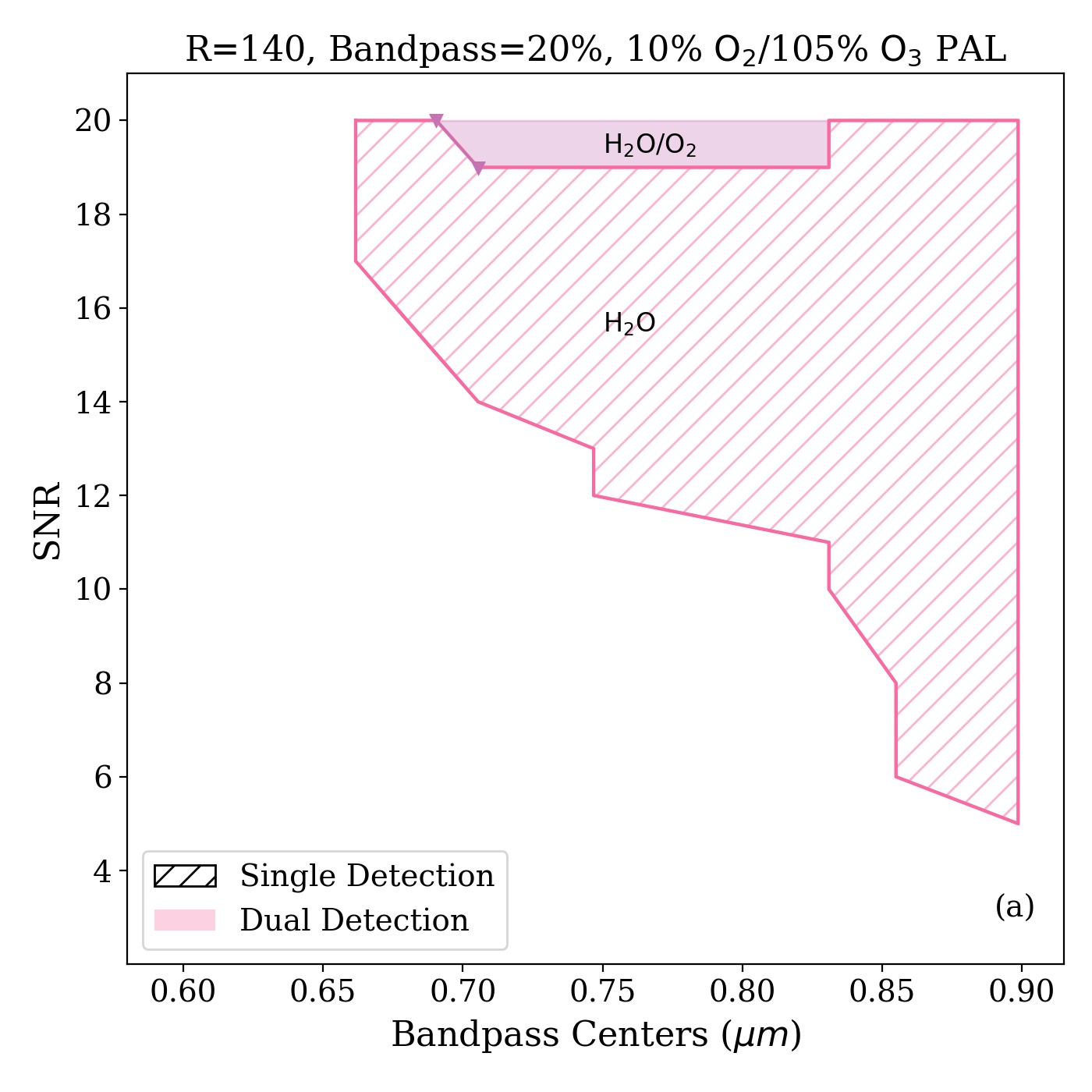}{0.334\textwidth}{}
          \fig{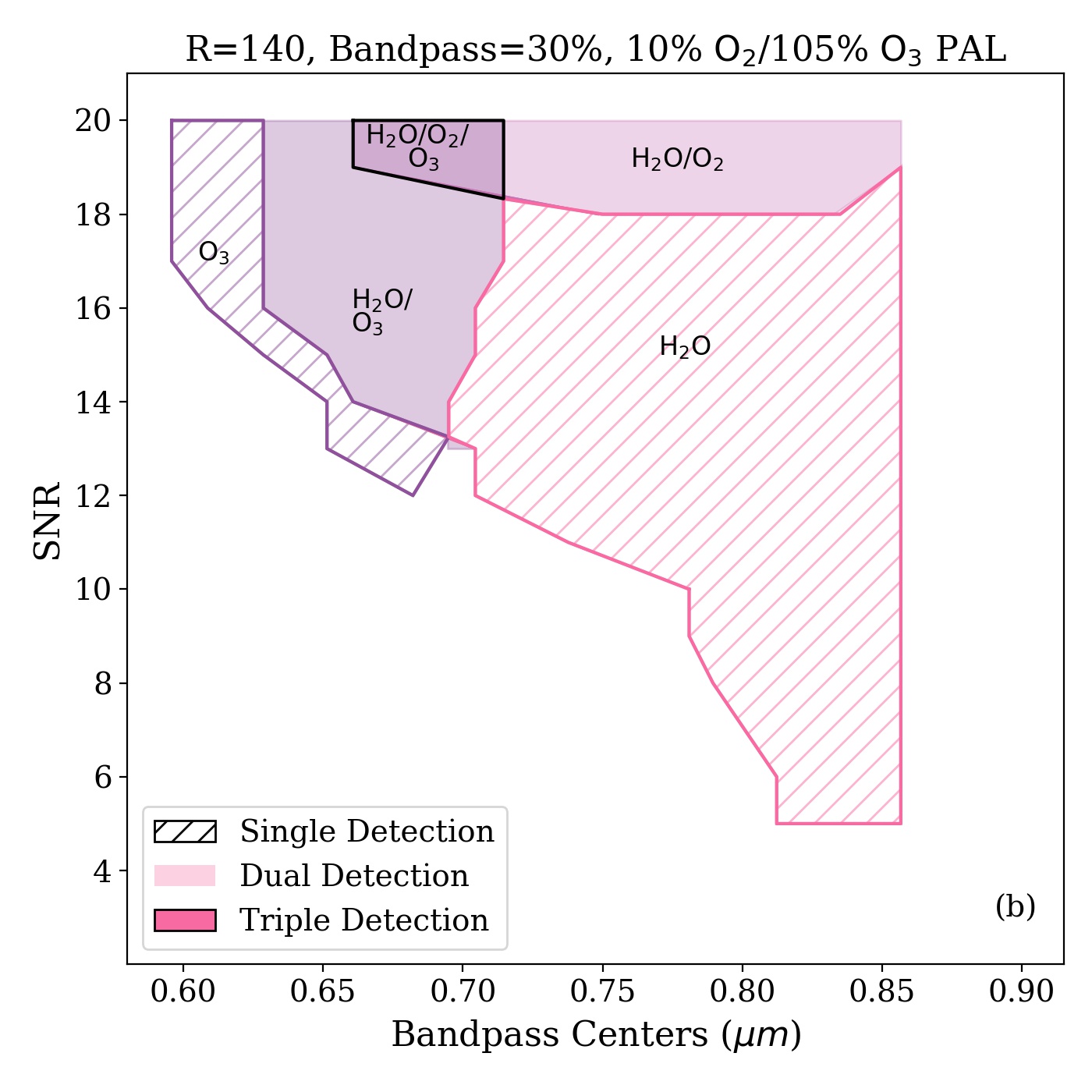}{0.334\textwidth}{}
          \fig{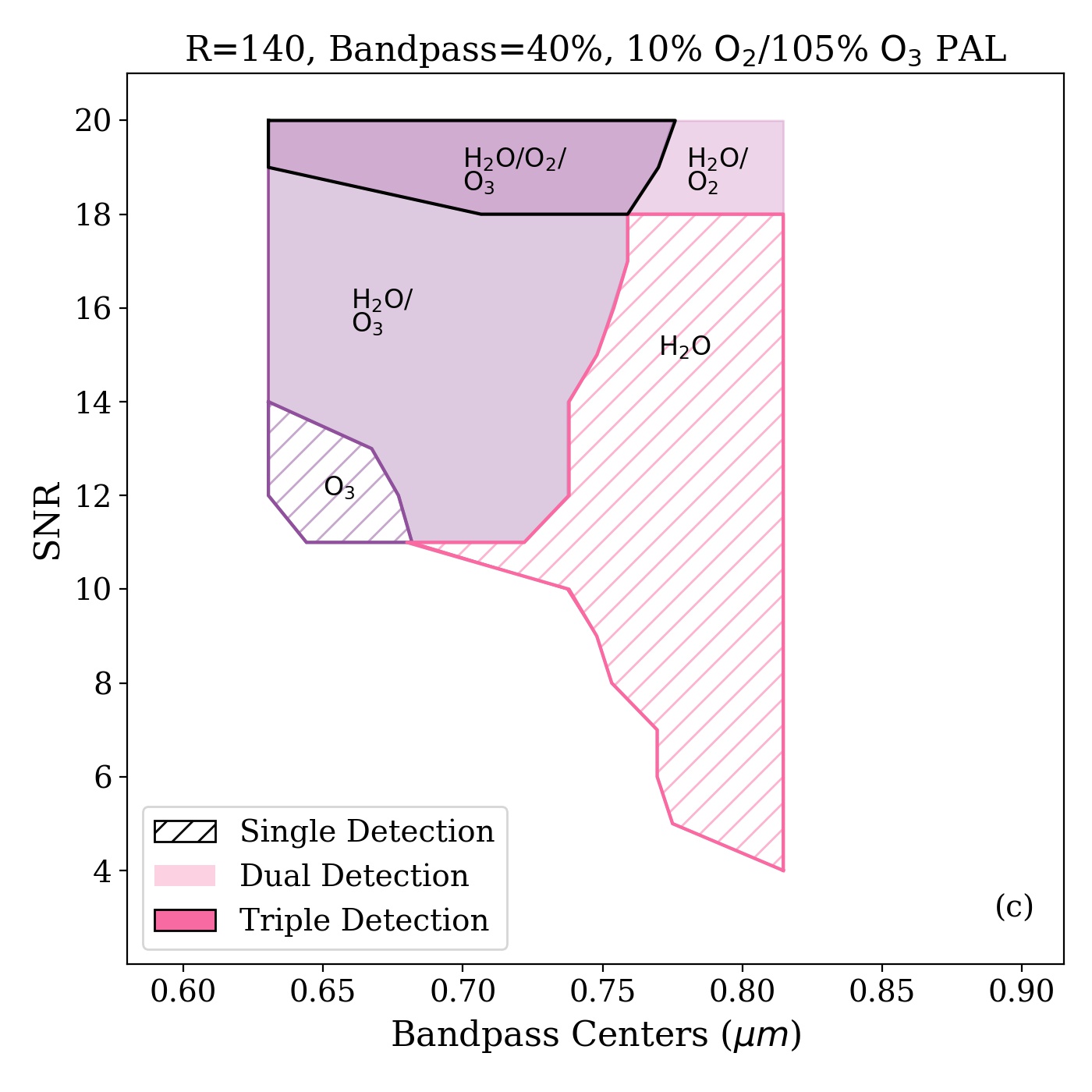}{0.334\textwidth}{}}
\vspace{-1cm}
\gridline{\fig{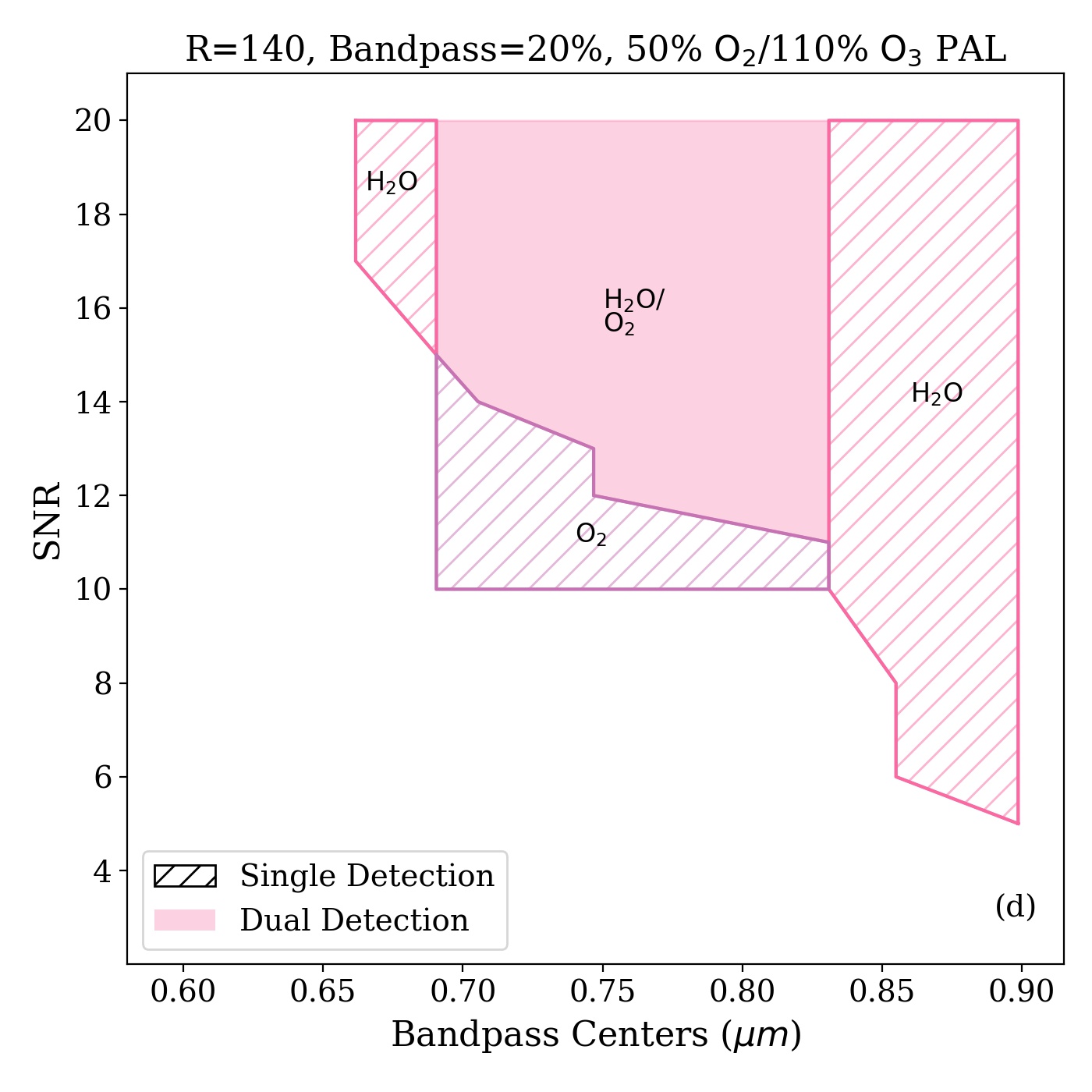}{0.334\textwidth}{}
          \fig{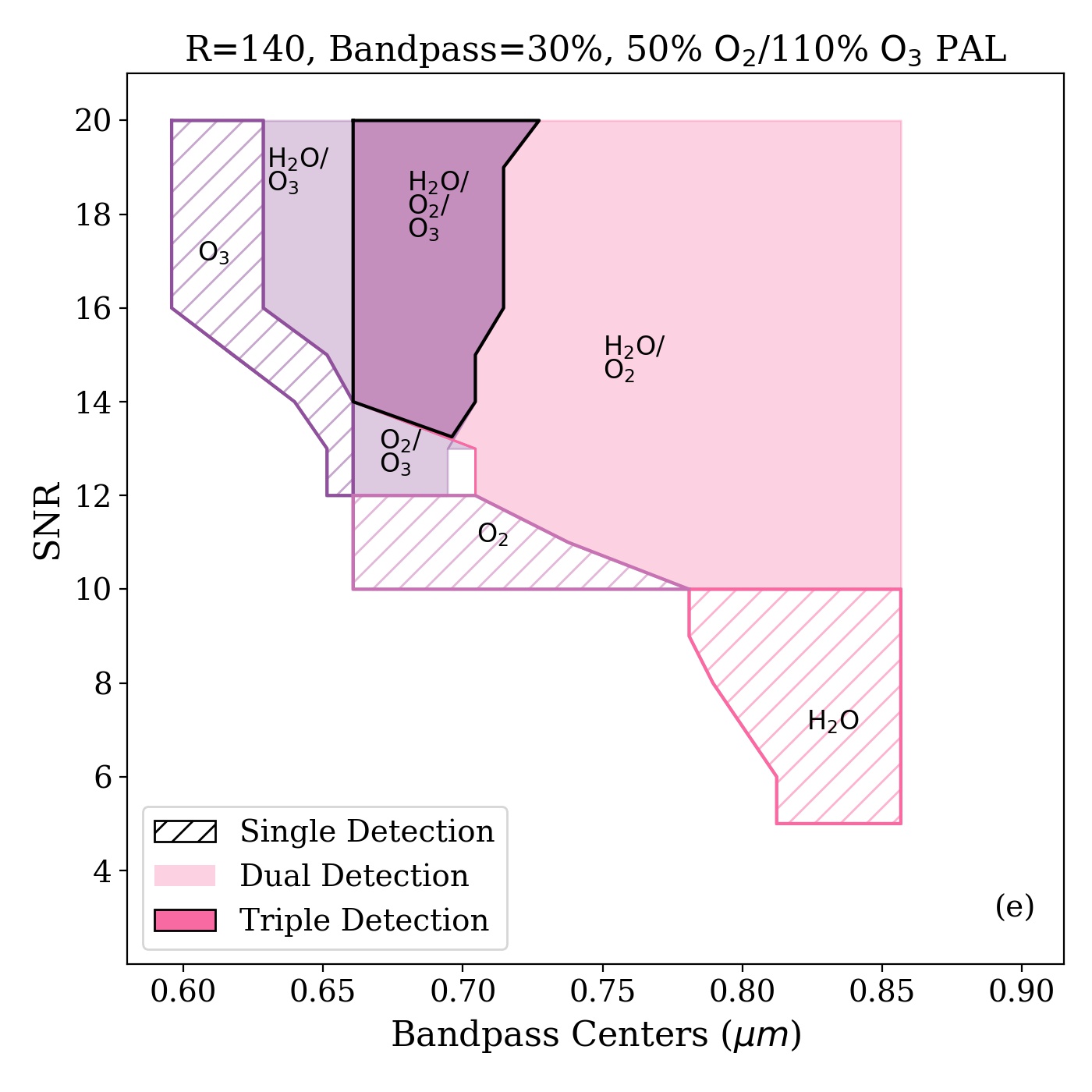}{0.334\textwidth}{}
          \fig{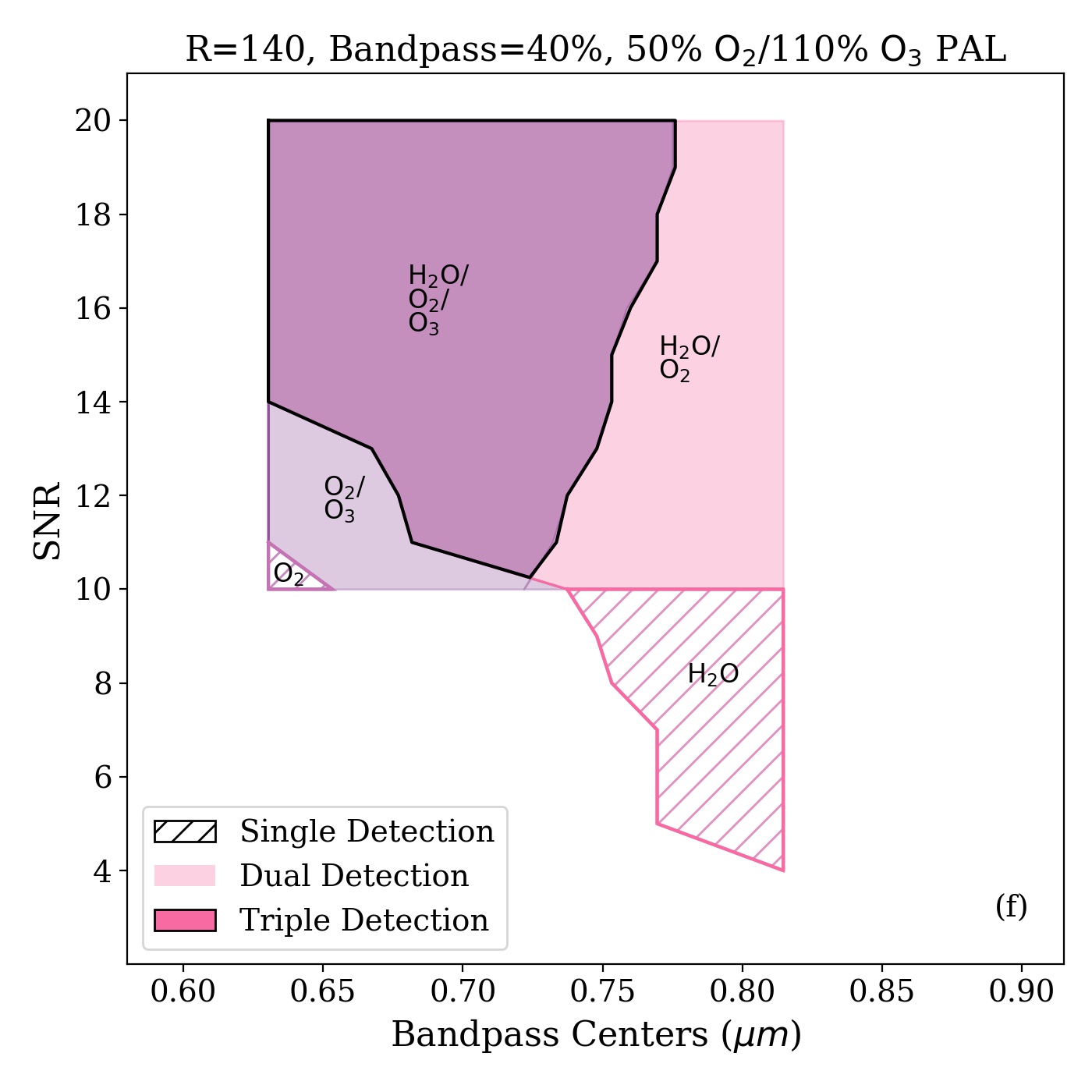}{0.334\textwidth}{}}
\vspace{-1cm}
\gridline{\fig{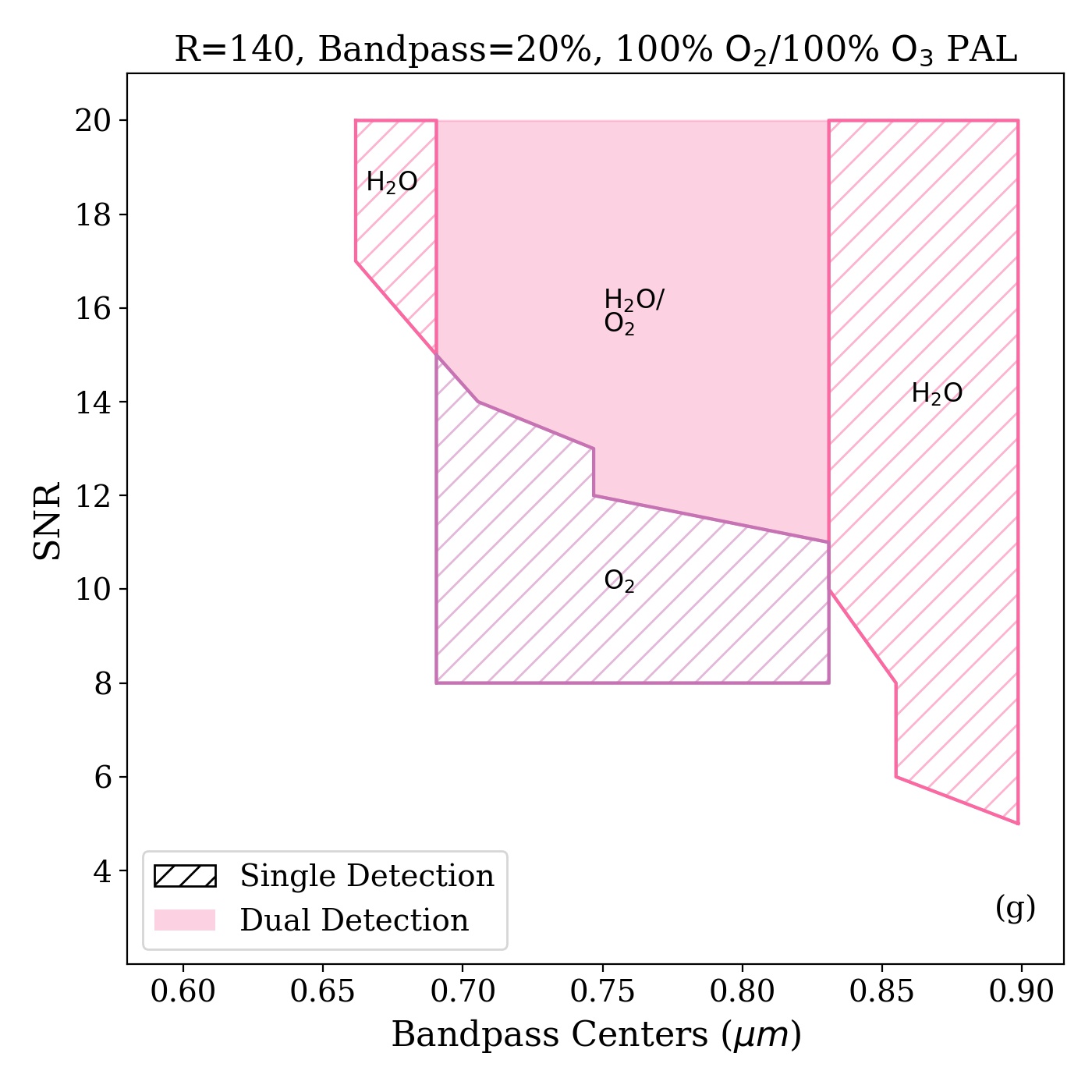}{0.334\textwidth}{}
          \fig{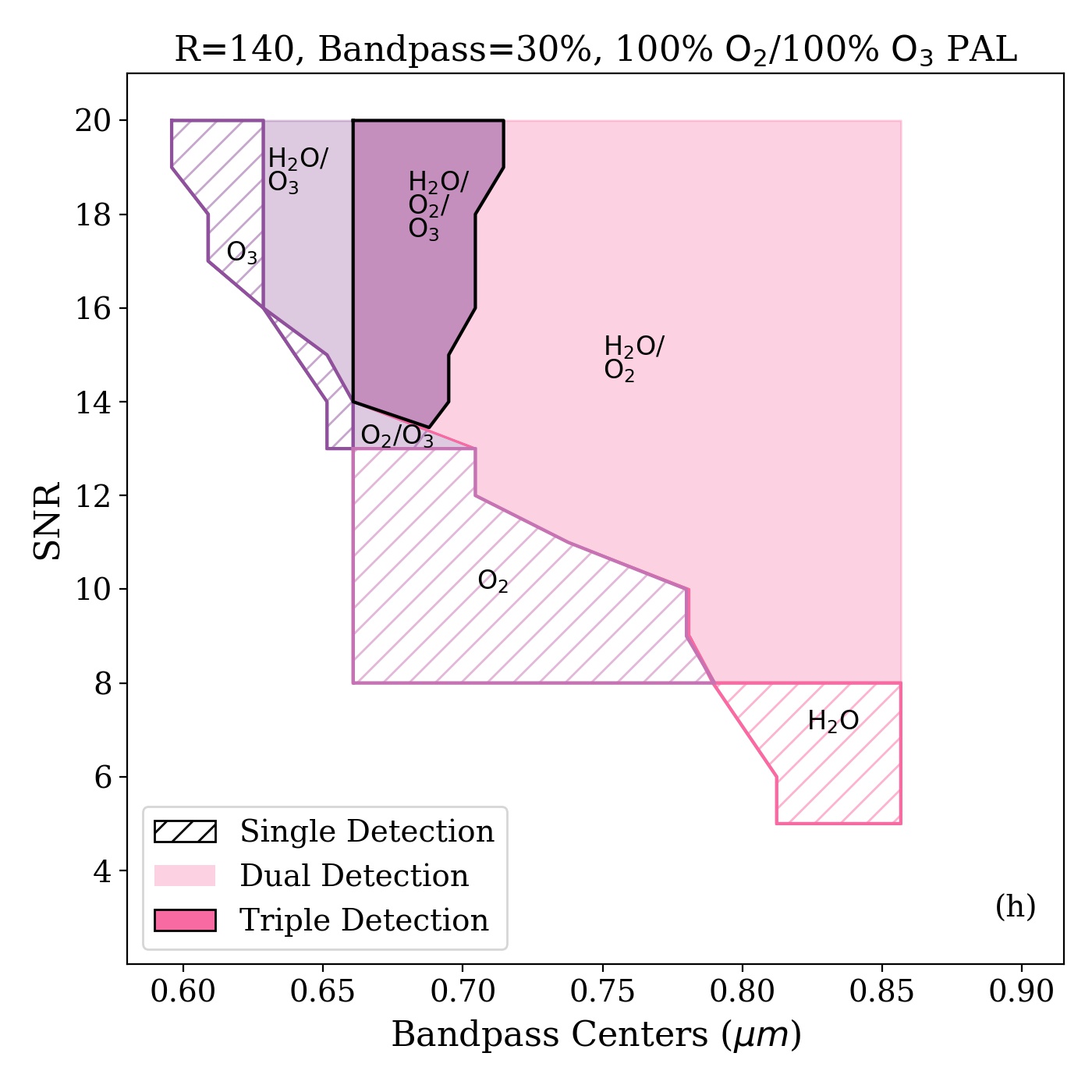}{0.334\textwidth}{}
          \fig{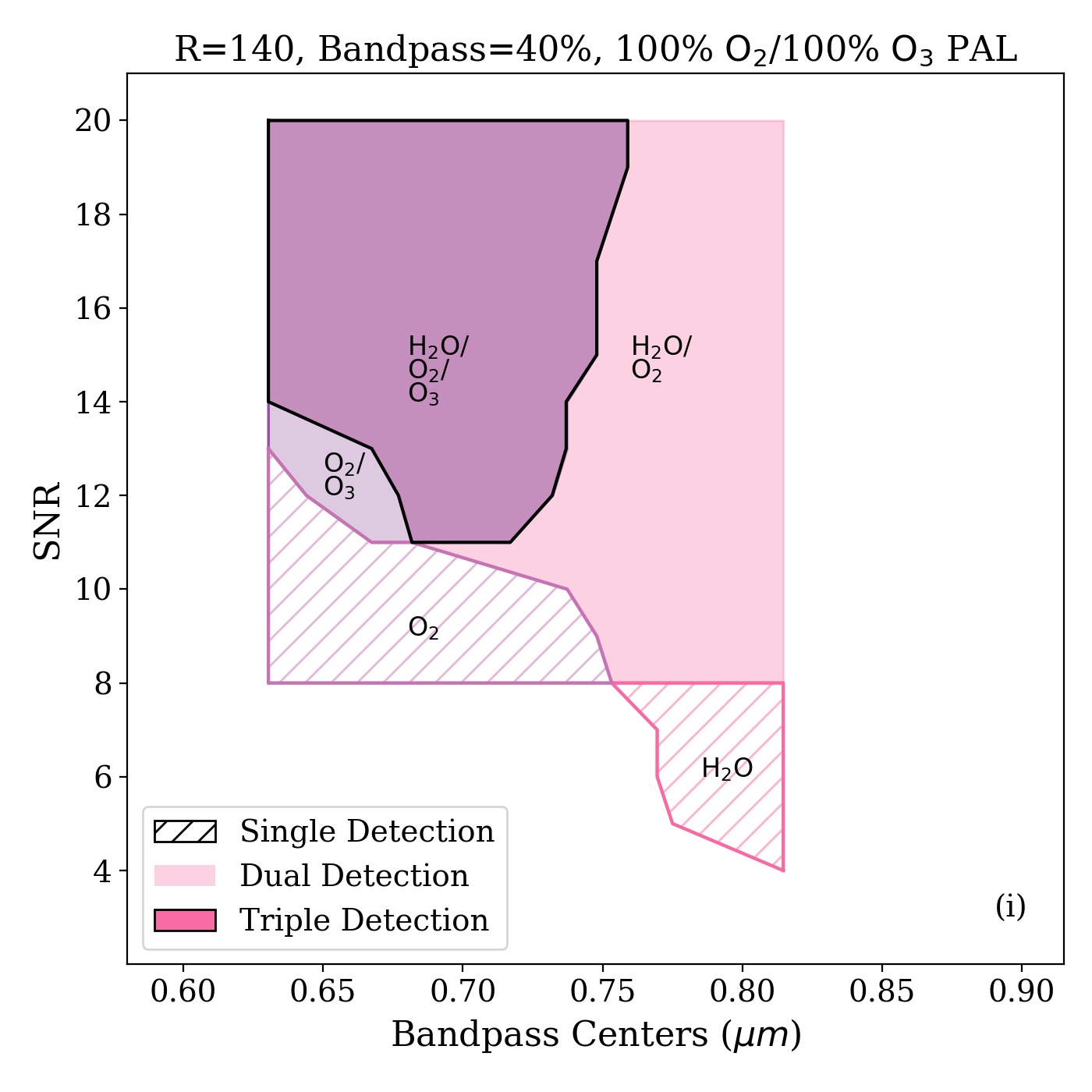}{0.334\textwidth}{}}
\vspace{-0.5cm}
\caption{Shown above are the regions of single, dual, and triple molecular detection as a function of wavelength. wavelength is presented on the x-axis, SNR is presented on the y axis. The hatched regions symbolize regions with single molecule detection, shaded regions symbolize only the overlapping areas between dual molecular strong detection zones, and the black outlined shaded regions symbolize the overlapping areas between triple molecular strong detection. We present the 10\% \ce{O2} PAL and 105\% \ce{O3} PAL case in the first row, the 50\% \ce{O2} PAL and 110\% \ce{O3} PAL case in the second row, and the 100\% \ce{O2} PAL and 100\% \ce{O3} PAL case in the last row. For all cases, we present a 20\%, 30\%, and 40\% bandpass width.}
\label{fig:o2o3testcomphatch}
\end{figure*}

As previously discussed, in our previous simulations we varied one molecular abundance (for \ce{O2}) while the other parameters were held fixed to modern Earth-like values, thus we did not explore the relationship between varying both \ce{O2} and \ce{O3} and the resultant change in detectability.  However, due to the photochemical relationship between the \ce{O2} abundance and the production of \ce{O3}, the abundances of both molecules would actually be linked in any atmospheric scenario. As found in \citet{kozakis22}, the relationship between \ce{O2} and \ce{O3} varies for model atmospheres depending on the type of host star, and the linearity of the relationship also appears to vary. When looking to hotter host stars (e.g., a G2 star), peak \ce{O3} abundance occurs at lower than modern Earth \ce{O2} abundances due to the \ce{O3} layer shifting in the atmosphere downwards to \ce{O2} levels due to \ce{O2} photochemical shielding.

This \ce{O2}/\ce{O3} relationship means that we can constrain the overall molecular oxygen abundance by detecting either of the species; \ce{O3} is in fact a highly sensitive tracer of lower abundances of \ce{O2}. In order to explore the impact of this, we examined several scenarios where we varied our parameter values as coupled parameters, i.e. 10\% PAL O2/105\% PAL O3, etc., consistent with the results for a Sun-like star from \citet{kozakis22} (as seen in Table~\ref{tab:atmosvals}). We present our detectability results for these abundance pairs in Figure~\ref{fig:o2o3testcomp}. In Figure~\ref{fig:o2o3testcomp}a-c, we present the \ce{O2} results with \ce{H2O} at 20\%, 30\%, and 40\% bandpasses at 10\% PAL, 50\% PAL, and 75\% PAL of \ce{O2}. We can see that at 10\% PAL, \ce{O2} detectability is unlikely, requiring a minimum SNR of 19 at a 20\% bandpass, and SNR of 18 at 30\% and 40\% bandpasses. However, between 50\% PAL, 75\% PAL, and the previously presented 100\% PAL of \ce{O2}, there are little difference, with slight variance in SNR (e.g. from requiring SNR of 8 for 100\% PAL \ce{O2} to requiring SNR of 10 for 50\% PAL \ce{O2}) and little change across bandpass width as in Figure~\ref{fig:shortest_detec}. In Figure~\ref{fig:o2o3testcomp}d-f, we present the \ce{O3} results with \ce{H2O} at 20\%, 30\%, and 40\% bandpasses at 105\% PAL and 110\% PAL of \ce{O3}. We can see that detectability does not shift significantly for \ce{O3}, as our values do not vary greatly from a modern-Earth like value. However, one notable difference is that there is a larger range for strong \ce{O3} detection at both 105\% and 110\% \ce{O3} PAL with bandpass width 20\% than the same bandpass width with 100\% \ce{O3} PAL as in Figure~\ref{fig:shortest_detec}. We also see that \ce{O3} is detectable at an SNR of 17 rather than 19 as in the modern-Earth like results for a 20\% bandpass. The detectability range of \ce{O3} increases with increasing bandpass width, as expected following Figure~\ref{fig:shortest_detec}.

Following this, we analyze the limiting molecule in double or triple molecular detection in Figure~\ref{fig:o2o3testcomphatch}. We present 10\% \ce{O2}/105\% \ce{O3} PAL in Figure~\ref{fig:o2o3testcomphatch}a-c. We can see that detection of \ce{O2} dually or triply with \ce{H2O} or \ce{O3} is difficult. At a 20\% bandpass, \ce{H2O} and \ce{O2} are dual detectable only at an SNR $\ge19$. At a 30\% bandpass, triple \ce{H2O}/\ce{O2}/\ce{O3} detection is possible at an SNR $\ge18$ in a very narrow wavelength range (approximately 0.66 to 0.7 {\microns}), and dual \ce{H2O}/\ce{O2} detection is also possible at an SNR $\ge18$, while dual \ce{H2O}/\ce{O3} detection has a wider range of detectability, from SNR $\ge14$, and from approximately 0.63 to 0.7 {\microns}. At a bandpass of 40\%, the range for triple detection increases in wavelength, to approximately 0.63 to 0.76 {\microns}. The range for dual \ce{H2O}/\ce{O3} detection also increases, starting from SNR $\ge11$ and from 0.63 to 0.75 {\microns}. At all bandpasses, \ce{H2O} has a large range in wavelength and SNR of single detectability thus it is highly likely to strongly detect \ce{H2O}. In all bandpasses, \ce{O2} is the limiting molecule for dual or triple detection in both SNR and wavelength. In terms of dual \ce{H2O}/\ce{O3} detection, \ce{O3} is the limiting molecule in wavelength, but \ce{H2O} is the limiting molecule in SNR.

We present 50\% \ce{O2}/110\% \ce{O3} PAL in Figure~\ref{fig:o2o3testcomphatch}d-f. We can see that, as in the modern Earth case, dual detection of \ce{H2O} and \ce{O2} is always possible at all bandpasses. The same is not true for \ce{H2O} and \ce{O3} however, with dual strong detection possible at only a single point in the 20\% bandpass, and with a wider range possible at 30\%. At 40\%, there is no dual detection of \ce{H2O} and \ce{O3}, however there is a large range (approximately 0.12 {\microns} in width) wherein triple detection is possible from SNR $\ge11$. At 30\% and 40\% bandpasses there are also areas where dual detection is possible with \ce{O2} and \ce{O3} at mid-high SNR and short wavelengths. With a 20\% bandpass, \ce{H2O} is the limiting molecule for dual \ce{H2O}/\ce{O2} detection in both SNR and wavelength. At a 30\% bandpass, \ce{H2O} is still the limiting molecule for dual \ce{H2O}/\ce{O2} detection and triple \ce{H2O}/\ce{O2}/\ce{O3} detection in SNR, however \ce{O3} is the limiting molecule in wavelength. When looking at dual \ce{H2O}/\ce{O3} detections, \ce{H2O} is the limiting molecule in SNR and wavelength, mimicking the dual \ce{H2O}/\ce{O2} detection. At a 40\% bandpass, \ce{O3} is the limiting molecule in wavelength for a triple \ce{H2O}/\ce{O2}/\ce{O3} detection, and \ce{H2O} is the limiting molecule in SNR for a triple \ce{H2O}/\ce{O2}/\ce{O3} detection. We then have the same \ce{H2O} limiting for dual \ce{H2O}/\ce{O2} detection in SNR, although there is no limiting molecule for dual detection in wavelength since the final bandpasses have detections for both \ce{H2O} and \ce{O2}.

We present 100\% \ce{O2}/100\% \ce{O3} PAL in Figure~\ref{fig:o2o3testcomphatch}g-i. In the 20\% bandpass, we see extremely similar results to Figure~\ref{fig:o2o3testcomphatch}d, with the notable difference that \ce{O2} detection is possible down to an SNR of 8, compared to an SNR of 10. In the 30\% bandpass, we again see almost identical results to the 50\% \ce{O2}/110\% \ce{O3} PAL case shown in Figure~\ref{fig:o2o3testcomphatch}e. The differences here are once again that \ce{O2} is detectable down to an SNR of 8, but we also notice that \ce{O3} has a smaller region of single detectablity, which is also seen in the smaller region of dual \ce{O2}/\ce{O3} detection. As a result of the lower SNR requirements for \ce{O2} detectability, this results in a large region of dual \ce{H2O}/\ce{O2} detection. In the 40\% bandpass, once again the results mimic the 50\% \ce{O2}/110\% \ce{O3} PAL counterpoint case shown in  Figure~\ref{fig:o2o3testcomphatch}f. The differences shown are a smaller region of triple \ce{H2O}/\ce{O2}/\ce{O3} detection and dual \ce{O2}/\ce{O3}, resulting in larger single \ce{O2} and dual \ce{H2O}/\ce{O2} detection regions. This also results in a smaller single \ce{H2O} detection region. In these cases, the limiting molecules are the same as above: At a 20\% bandpass, \ce{H2O} is the limiting molecule for dual \ce{H2O}/\ce{O2} detection in both SNR and wavelength. At a 30\% bandpass, \ce{H2O} is the limiting molecule for dual \ce{H2O}/\ce{O2} detection and triple \ce{H2O}/\ce{O2}/\ce{O3} detection in SNR, however \ce{O3} is the limiting molecule in wavelength. At a 40\% bandpass, \ce{O3} is the limiting molecule in wavelength and SNR for a triple \ce{H2O}/\ce{O2}/\ce{O3} detection.

\section{Discussion}
\label{sec:disc}

The primary utility of the retrieval results presented here is to help understand how the impact of \ce{O2} and \ce{O3} abundance affects the optimal strategy for detecting the presence of an atmosphere with water and oxygen with spectroscopic bandpasses in the visible-light spectral region. Here we discuss the major conclusions from this work, as well as some of the limitations to the study that may impact these conclusions.

The most important result is the role that the width of the bandpass plays. When examining Figures~\ref{fig:epoch_o2} and \ref{fig:epoch_o3}, we can see the notable difference in how a widening bandpass changes the detectability of molecules across abundance and type; where \ce{O2} detectability does not vary significantly as a function of bandpass across abundance values, \ce{O3} changes significantly as the bandpass widens. This is likely due to the fact that the \ce{O2} feature is narrow and deep, thus making it easily detectable at high abundance values, but easily captured in a smaller bandpass width, whereas \ce{O3} has a very broad feature thus the widening of the bandpass allows for a stronger continuum and significant change in detection as a function of bandpass. \ce{O3} also does not have any strong features in the 0.515 -- 1 {\microns} region, leading to inherent difficulty in detection in this range. However, if we are able to observe the strong \ce{O3} band at 0.36 {\microns}, detection would become stronger at lower SNR data. Recent works, such as by \citet{damiano23}, have investigated this principle and come to the conclusion that even with impacts of possible confusion due to \ce{SO2} and other species, \ce{O3} detection is indeed easier at lower SNR and lower resolution data. \citet{damiano23} found that with spectral resolution of 7, and SNR = 10, \ce{O3} can be detected in the UV at sub-PAL VMRs; they also note that observing additional signatures of habitability in the NIR region is crucial to interpreting \ce{O3} detections at UV and VIS wavelengths.We will examine these questions further in future work addressing the complete UV, VIS and NIR wavelength regions.

A wider bandpass also better enables detections of one or more molecules with a single bandpass - in particular, the detection of \ce{O3} with other molecules.  In Figure~\ref{fig:shortest_detec}, we can see that for modern Earth abundances a bandpass width of $\ge30\%$ would allow for a detection of \ce{H2O} at slightly shorter wavelengths (long-wavelength cutoff of $\sim$0.95 {\microns} versus 0.99 {\microns} for a 20\% bandpass) and also a joint measurement of a strong \ce{O2} detection and at least a weak \ce{O3} detection with SNR = 8-9 versus only \ce{O2} (which could help to limit false positives and motivate additional observations). Furthermore, Figure~\ref{fig:o2o3testcomphatch} shows that a 40\% bandpass width enables a significant improvement in the ability to detect all three species with a single bandpass at around 0.72 {\microns}; this could either be used alone as a first reconnaissance bandpass or could be used as the follow-up to a longer-wavelength search for \ce{H2O} alone at low SNR. The optimal choice between these two options depends significantly on whether a planet is detectable at longer wavelengths (due to IWA) and whether the exposure time required for longer-wavelength measurements is significantly impacted by instrument sensitivity; we leave these instrument- and target-dependent considerations to future work.

The second important result is that the abundance versus SNR results in Figures~\ref{fig:epoch_o2} and \ref{fig:epoch_o3} show a non-linear increase in the required SNR at smaller abundances. Combined with the fact that SNR is linearly dependent on the square of exposure time, this suggests that our exposure time will be extremely sensitive to the limiting abundance that must be detectable.  Figure~\ref{fig:o2o3testcomphatch} further demonstrates this, showing that when \ce{O2} drops from 50\% PAL (12\% VMR) to 10\% PAL (2\% VMR), it essentially becomes undetectable. This motivates a deeper analysis of whether detecting \ce{O2} at visible wavelength should be a high priority in the progression of measurements if there is a high likelihood of a non-detection for even a moderate abundance, compared with a measurement of more sensitive markers of atmospheric abundance in the UV or NIR.

We note that high \ce{O3} abundances are challenging to model, in that it takes a very small amount to completely overwhelm the spectrum. Spectra with high \ce{O3} have continua near zero, which also causes the error value to increase substantially, thus high values of \ce{O3} must be handled with care. At high abundances, the error value increase can lead to poorly constrained retrievals using our grid. \citet{susemiehl23} found that this occurs at high \ce{O3} ($log_{10}\ce{O3}>-5$) concurrently occurring with low $\mathrm{P_{0}}$ ($log_{10}\mathrm{P_{0}}<-1.7$) using our grid. Our simulations go no higher than $log_{10}\ce{O3}\leq-6$ and therefore avoid this degeneracy and resulting poorly constrained retrievals.  

As discussed in BARBIE1, there have been similar prior works that investigated the relationship between SNR and detectability, specifically \citet{feng18} and \citet{damiano22}. We explore the differences in retrieval parameterization structure in BARBIE1, however even with the differences in techniques and analysis, our results for \ce{O2} and \ce{O3} detectability are in agreement. An SNR of 10 at R = 140 is sufficient to firmly constrain the abundances for an Earth-twin atmosphere with \ce{O2} and \ce{O3}. Our work also explores the influence of varying bandpass width on molecular detection, which varies from prior work, and thus presents a new analysis of detection possibilities.

\section{Conclusions \& Future Work}
\label{sec:conc}

To summarize, by investigating bandpass width in tandem with SNR, we can see that detectability is intrinsically linked to both factors as an additional function of molecular type. The ability to properly prioritize and select the best combination of parameters can drive efficient observing practices. By understanding the SNR requirements for strong detection of \ce{O2} and \ce{O3} and also understanding which bandpass width could result in simultaneous strong detection of \ce{O2}, \ce{O3}, and \ce{H2O}, we can properly prioritize the best combination of bandpass and required SNR for detection, thereby informing the best options for instrument design trades and observing procedure. \ce{O3} is most easily observable using a 30 or 40\% bandpass width at shorter wavelengths with mid-low SNR data. With a 20\% bandpass width, \ce{O3} is difficult to detect, requiring SNR $\ge19$ at modern-Earth values. \ce{O2} is not significantly affected by bandpass width, and consistently requires an SNR of $\ge8$ to be strongly detected at modern-Earth values at 0.76 {\microns} and shorter. It is not detectable at SNR $\le20$ at Proterozoic era abundances. Since the coronagraph and instrument capabilities may dictate that the best observation occurs at shorter wavelengths, \ce{O2} and \ce{O3} are well within the most optimized instrument capabilities. \ce{O2} and \ce{O3} have strong geochemical markers that provide a depth of knowledge to potential Earth observations, and a heightened ability to constrain the observed Earth-like epoch. 

By also investigating coupled atmospheric abundances of \ce{O2} and \ce{O3}, we can study how detectability of these molecules vary with the other parameter. Allowing parameters to vary with each other following leading coupled photochemistry and atmospheric models, we can prioritize the optimal bandpass width for more realistic exo-Earth simulations. 

In future work, we plan to build a new spectral model grid that includes a wider range of molecular species, and we will extend to shorter and longer wavelengths than the visible range. Specifically, we will cover the same wavelengths as the proposed coronagraph instrument for the Habitable Worlds Observatory \citep{roser22}, including UV, Optical, and NIR. This will allow us to study more biosignatures and molecular species of interest, and represent the physical chemistries more accurately. This will allow us to expand our simulations of possible observations and establish best observational practices for exoEarth observations using next generation telescopes. We will also develop a PSG module to display all detectability information in a publicly accessible format. We note that SNR results may be subject to minor changes following grid reconstruction due to PSG radiative transfer upgrades. We will also include a full exoplanet yield calculation using the data contained within BARBIE1 and BARBIE2 to give an educated baseline for detection of biomarkers. The interpolation metric used in this work is a trilinear interpolation scheme which can be used on 3D grids as in this work, but in future works we will investigate if another interpolation metric, such as inverse distance weighted interpolation or multiplicative weight interpolation, can minimize interpolation error in grid-based retrievals. 

We will also conduct photochemically self-consistent studies in order to inform and broaden the results for future retrieval studies using grids. The expansion of grid parameter space will allow for the possibility of chemically consistent modeling, combined with input from updated photochemical models to ensure self-consistent atmospheric gas compositions with retrievals. This will be particularly important for considerations of gas detections for planets around different star types, given the impact star type has on the photochemistry of planetary atmospheres \citep{segura2005, segura2010, france12}.

N. L. gratefully acknowledges financial support from an NSF GRFP. N.L. gratefully acknowledges Dr. Joesph Weingartner for his support and editing. N. L. also gratefully acknowledges Greta Gerwig, Margot Robbie, Ryan Gosling, Emma Mackey, and Mattel Inc.{\texttrademark} for Barbie (doll, movie, and concept), for which this project is named after. This Barbie is an astrophysicist! The authors also gratefully acknowledge conversation with Dr. Chris Stark regarding exoEarth yields and instrument design. The authors would like to thank the Sellers Exoplanet Environments Collaboration (SEEC) and ExoSpec teams at NASA's Goddard Space Flight Center for their consistent support. MDH was supported by an appointment to the NASA Postdoctoral Program at the NASA Goddard Space Flight Center, administered by Oak Ridge Associated Universities under contract with NASA.



\bibliographystyle{aasjournal}
\bibliography{main}{}

\end{document}